\documentclass[letterpaper,onecolumn,11pt]{IEEEtran}
\usepackage[cmex10]{amsmath}
\usepackage{amsfonts}
\usepackage{amssymb}
\usepackage{amsthm}
\usepackage{bbm}
\usepackage{amscd}
\usepackage[pdftex]{graphicx}
\usepackage{epsfig}
\usepackage{mathrsfs}
\usepackage{afterpage}
\usepackage{cite}
\usepackage{enumerate}
\usepackage{url}
\newcommand{\nn}{\nonumber\\}
\newcommand{\C}{\mathbf{C}}
\newcommand{\X}{\mathcal{X}}
\newcommand{\PP}{\mathcal{P}}
\newcommand{\Y}{\mathcal{Y}}

\newcommand{\E}{\mathbb{E}}
\newcommand{\R}{\mathbb{R}}
\newcommand{\V}{\text{Var}}
\newcommand{\Vm}{V_{\min}}
\newcommand{\F}{\mathcal{F}}
\newcommand{\Hc}{\mathcal{H}}

\newcommand{\G}{\mathcal{G}}

\newcommand{\ii}{\mathfrak{i}}

\newcommand{\A}{\mathbf{A}}
\newcommand{\B}{\mathbf{B}}
\newcommand{\W}{\mathbf{W}}
\newcommand{\N}{\mathbf{N}}
\newcommand{\de}{\mathrel{\mathop:}=}
\newcommand{\1}{\mathbf{1}}
\newcommand{\as}{\text{a.s.}}
\newcommand{\nmx}{\nu_{\text{max}}}
\newcommand{\nmn}{\nu_{\text{min}}}
\newcommand{\imx}{\mathfrak{i}_{\text{max}}}
\newcommand{\Vmx}{V_{\max}}
\newcommand{\cd}{\mathfrak{C}}

\newtheorem{Lemma}{Lemma}

\newcommand{\eqdef} {\mathrel{\mathop:}=}
\newtheorem{definition}{Definition}
\newtheorem{remark}{Remark}
\newtheorem{theorem}{Theorem}

\newtheorem{proposition}{Proposition}

\newcommand{\cX}{{\cal X}}
\newcommand{\cY}{{\cal Y}}

\newcommand{\cA}{{\cal A}}

\newcommand{\cP}{{\cal P}}
\newcommand{\cB}{{\cal B}}
\newcommand{\cS}{{\cal S}}

\newcommand{\cU}{{\cal U}}

\newcommand{\lm}{\Lambda}

\newcommand{\T}{\mathcal{T}}

\def\b0{\mathbf{0}}
\def\bX{\mathbf{X}}

\def\bY{\mathbf{Y}}

\def\bbR{\mathbb{R}}
\def\bbZ{\mathbb{Z}}
\def\bx{\mathbf{x}}
\def\by{\mathbf{y}}

\def\mE{\mathbb{E}}
\def\me{\textnormal{e}}

\def\mo{\textnormal{o}}
\def\mI{I}
\def\mC{C}

\def\mV{V}
\def\b1{{\mathbf{1}}}
\def\mP{\textnormal{P}}

\newcommand{\Bf}{\mathbf{f}}

\newcommand{\dtv}{d_{\text{TV}}}

\def\eor{\hskip 5 pt $\Diamond$}
\newcommand*{\Qed}{\hskip 5 pt $\IEEEQEDopen$}
\renewcommand*{\IEEEQED}{\IEEEQEDopen}

\begin{document}
\title{A New Method for Employing Feedback to Improve Coding Performance}
\author{Aaron B.~Wagner, Nirmal V.~Shende, and Y\"{u}cel Altu\u{g}}
\maketitle
\begin{abstract}
We introduce a novel mechanism, called timid/bold coding,
by which feedback can be used to
improve coding performance. For a certain class of DMCs,
called \emph{compound-dispersion channels},
we show that timid/bold coding allows for an improved second-order
coding rate compared with coding without feedback.
For DMCs that are not compound dispersion, we show 
that feedback does not improve the second-order coding rate.
Thus we completely determine the class of DMCs for which
feedback improves the second-order coding rate. An upper bound
on the second-order coding rate is provided for compound-dispersion
DMCs. We also show that feedback does not improve the second-order 
coding rate for very noisy DMCs. The main results are obtained
by relating feedback codes to certain controlled diffusions.
\end{abstract}

\section{Introduction}
\label{sec:intro}
Consider the canonical communication model consisting of a
single encoder sending bits to a single decoder over a discrete
memoryless channel (DMC). We assume the alphabets are finite, the
channel law is completely known, and the transmission rate is fixed,
i.e., the decoding of the entire message must occur at a 
prespecified time.

In practice, point-to-point communication links are usually paired
with a feedback link from the decoder to the encoder, which can
communicate messages in the reverse direction but can also be used
to facilitate communication along the forward link. Although
such feedback links are common in practice, it is not well understood
theoretically how they can be most effectively used. We 
consider how unfettered use of a 
perfect feedback link can improve asymptotic coding performance across 
the forward channel.
It is well-known that feedback does not improve the capacity of
a DMC~\cite{Shannon:Zero}. We shall consider how feedback can used to 
improve the more-refined \emph{second-order coding rate} of the 
channel (see Def.~\ref{def:SOCR} to follow).

\emph{A priori}, it is not clear that feedback improves
the second-order coding rate at all. Indeed, none of the
mechanisms by which feedback is known to improve coding
performance obtains for the setup under study. The channel
has no memory, so feedback cannot be used to anticipate
future channel disturbances (as in, e.g.,~\cite{Kim:Feedback}). The 
channel law is known, so feedback is not useful for learning 
the channel statistics (as in, e.g.,~\cite{Tchamkerten:Unknown}).
The blocklength is fixed, so feedback does not allow the 
code to outwait unfavorable noise realizations 
(cf.~\cite{Burnashev:Feedback}). There is no cost constraint,
so the encoder cannot use feedback to opportunistically
consume resources (cf.~\cite{Schalkwijk:Feedback:I,Schalkwijk:Feedback:II}).
Since the second-order coding
rate focuses on a ``high-rate'' regime, the increase in
the effective minimum distance of the code afforded by feedback is
not useful (cf.~\cite{Berlekamp:PHD}). Since the channel is
point-to-point, none of the various ways that feedback
can enable coordination in networks (e.g.,~\cite{Gaarder:MAC})
can be applied.  Indeed, a negative result
is available showing that feedback does not increase the
second-order coding rate for DMCs satisfying a certain
symmetry condition~\cite[Theorem 15]{PolyanskiyFB}.

We introduce a novel mechanism by which feedback can improve
coding performance for some DMCs, even when the coding
is high-rate and fixed-blocklength and the channel is known
and memoryless. The idea is the following.
Suppose a player may flip one of two fair coins in each of
$n$ rounds. If the player chooses to flip the first (resp. second)
coin, then she wins \$1 (resp.\ \$2) with probability half and loses
\$1 (resp.\ \$2) with probability half. We assume that each flip of
each coin is independent of everything else and that the initial
wealth is $w\sqrt{n}$ with $w > 0$. The player wins the overall game
if her wealth after $n$ rounds is positive. How should the player
decide which coin to flip in a given round in order to maximize
her chance of winning?
If the player is required to choose her strategy before
the start of the game, i.e., she is not allowed to update her
choice after seeing the previous flips, one can verify that
playing the first coin in all of the rounds is asymptotically
her best strategy. Indeed, under this strategy the central limit
theorem (CLT) implies that the probability of losing converges to
$\Phi( -w)$, where $\Phi$ is the distribution of the standard Gaussian
random variable. If she plays the second coin in all rounds, then
this probability is $\Phi( -w/2)$, which is worse. If she timeshares
the two coins, the probability will be in between. Essentially,
because she is expecting to win, she minimizes the probability
of losing by minimizing the variance of her wealth after round
$n$. Conversely, if she starts with $w < 0$, then she should
play the second coin for all time. Since she is expecting to 
lose, she minimizes the probability of losing by maximizing
the variance of her wealth after round $n$.

If the player can select the coin for each round using
knowledge of the outcomes of the previous rounds, 
then she can do better by utilizing both coins. Consider, for
simplicity, the scenario in which the player flips the first coin
for the first $n/2$ rounds and then selects one coin to flip for
all of the $n/2$ remaining rounds. A reasonable strategy is
the following: if the wealth after the first half is positive,
play ``timid," i.e., flip the coin that pays $\pm \$1$. Otherwise, play
``bold," i.e., flip the coin that pays $\pm \$2$. The justification is
that if her wealth is positive after $n/2$ rounds, then the player
is expecting to win, so she should minimize the variance of
her wealth after round $n$. If her wealth is negative after round $n/2$, then
she is expecting to lose, so she seeks to maximize the
variance after $n$ rounds. Another view is that if her wealth is
negative after round $n/2$, then she needs to have more wins
than loses during the second half in order to win overall; she
needs to be lucky. Quoting Cover-Thomas~\cite[p. 391]{cover-thomas}: ``If
luck is required to win, we might as well assume that we will
be lucky and play accordingly." Under the assumption that the
player will have more wins than loses, playing the coin that
pays $\pm \$2$ provides more wealth.

The connection to channel coding is provided by 
Lemmas~\ref{lemma:shannon:achievability:FB}
and~\ref{thm:gen_conv_code} in the Appendix, which relate the design 
of feedback
codes to the design of controllers for a particular controlled random 
walk.  For channels with multiple capacity-achieving input distributions 
that give rise to information densities with different variances,
which we call \emph{compound-dispersion} channels 
(see Definition~\ref{def:arm}), the controlled random walk that
arises through Lemmas~\ref{lemma:shannon:achievability:FB} 
and~\ref{thm:gen_conv_code} admits the timid/bold
play mechanism described above, and this in turn yields feedback
codes that asymptotically outperform the best non-feedback
codes. In channel-coding terms, the idea is that, with compound-dispersion
channels, the encoder can use codewords with symbols drawn from
multiple input distributions such that the mean rate of information 
conveyance across the channel is the same under all of these distributions 
(namely, the Shannon capacity), but the variance is different.
The encoder then monitors the progress of transmission via the
feedback link and uses a ``bold'' input distribution when
a decoding error is expected and a ``timid'' input distribution
when it is not. We call this \emph{timid/bold coding}.

Our course, it is desirable to update the strategy at each 
time during the block, instead of only at the halfway point.
This, however, comes at the expense of more technical arguments.
In particular, we  use convergence results for It\^{o} diffusion processes.
An inspiration   for this scheme is a result of McNamara on the optimal control of the diffusion coefficient  of a diffusion process~\cite{McNamara}.  Consider the following stochastic differential equation (SDE):
\[
\xi_t = \xi_0 +  \int_{0}^t\psi_s(\xi_s)\,dB_s \
\]
where $\xi_0$ is a constant, $0<\psi_s(x) \in[\psi_{\min}, \psi_{\max}]$ for all $s$ and $x$, and $\{B_t\}$ is a Brownian motion. If the goal is to maximize  ${P}\left({\xi}_1 \ge 0 \right)$ by choosing the function ${\psi}_s(\cdot)$, then McNamara shows that the bang-bang scheme
\begin{align}
\label{eq:bangbang}
{\psi}^{\text{opt}}(u)=
\begin{cases}
\psi_{\min} & u>0, \\
\psi_{\max} & u \le 0.
\end{cases}
\end{align}
is an optimal controller. If we view
this as a gambling problem then, in words,
the gambler should play maximally timid when she is expecting
to win and maximally bold when she is expecting to lose. 

McNamara~\cite{McNamara} notes that animals have been observed to follow
more-risky foraging strategies when near starvation and less-risky
strategies when food reserves are high. Similar behavior
is observed in sports, where, e.g., a hockey team will leave 
its goal unprotected in order to field an extra offensive
player if it is losing late in the game.  In the context of feedback 
communication, we show that
timid/bold coding improves the second-order
coding rate compared with the best non-feedback code for all 
compound-dispersion channels. We also show a matching converse result,
namely that feedback does not improve the second-order coding rate
of simple (i.e., non-compound) dispersion channels, improving
upon~\cite[Theorem 15]{PolyanskiyFB}. Thus, timid/bold 
coding provides
a second-order coding rate improvement whenever such an improvement
is possible\footnote{We assume throughout that the channel satisfies
$\Vm > 0$ as explained in the next section.}.
The converse is obtained by using the code modification technique 
of Fong and Tan~\cite{Fong-Tan} along with a ``Berry-Esseen''-type 
martingale CLT and large deviations results for 
martingales. In particular, this settles the problem of determining 
whether feedback improves the second-order coding rate for a given DMC.

For compound-dispersion channels, it is not clear if timid/bold
coding is an optimal feedback signaling scheme. To
shed some light on this question,
we provide the first nontrivial impossibility result for
the second-order coding rate of feedback communication 
over DMCs. The technical challenge in proving such a result
is that standard martingale central limit theorems do not provide useful bounds. Instead, we obtain the result using tools from stochastic calculus, namely, martingale embeddings, change-of-time methods, and McNamara's solution to the above-mentioned SDE. The bound on the second-order coding rate that we obtain
is functionally identical to the second-order coding rate achieved
by timid/bold coding, although evaluated at different channel 
parameters. The two
bounds coincide for some channels but not in general.

Finally, we show that feedback does not improve the  second-order coding rate for a class of DMCs called \emph{very noisy channels (VNCs)}. Reiffen~\cite{Reiffen} introduced VNCs to model physical channels that operate at a very low signal-to-noise ratio.\footnote{The VNCs introduced by Reiffen are called Class I VNCs by Majani~\cite{Majani}, where he also defined Class II VNCs. In this paper, we focus on Class I VNCs and refer to them simply as VNCs.}
VNCs are useful for modeling channels in which a  resource (such as power) is spread over many degrees of freedom (such as bandwidth)~\cite{Lapidoth}. 
We show that DMCs behave as simple-dispersion channels in the very noisy limit, and that feedback does not improve the second-order rate in this asymptotic regime. However, since DMCs only satisfy the simple-dispersion property in the limit, our converse for simple-dispersion channels is not directly applicable. Hence, we use a different proof technique.

The balance of the paper is organized as follows. The next section
describes the problem formulation more precisely and states all
five of our results. The remaining five sections then
provide the proofs of these five theorems in order. As described
earlier, the Appendix provides two lemmas that relate the design 
of feedback codes to the design of controllers for controlled random walks.
Although these lemmas have strong precedents in the literature,
the connection between feedback signaling and controlled random
walks seems to be novel.

\section{Notation, definitions and statement of the results}
\label{sec:notation-statement}
\subsection{Notation}
\label{ssec:notation}
$\bbR, \bbR^{+}$, $\bbR^-$ and $\bbR_{+}$ denote the set of real, positive real, negative real and non-negative real numbers, respectively. $\bbZ^+$ denotes the set of positive integers. We assume the input alphabet, $\mathcal{X}$, and the output alphabet, $\mathcal{Y}$, of the channel are finite. For a finite set $\cA$, $\cP(\cA)$ denotes the set of all probability measures on $\cA$. Similarly, for two finite sets $\cA$ and $\cB$, $\cP(\cB|\cA)$ denotes the set of all stochastic matrices from $\cA$ to $\cB$. Given any $P \in \cP(\cA)$ and
$W \in \cP(\cB|\cA)$, $P \circ W$ denotes the distribution
$$
(P \circ W)(a,b) = P(a)W(b|a).
$$
Given any $P \in \cP(\cA)$, $\cS(P) \eqdef \{ a \in \cA \, : \, P(a) >0\}$. $\Phi(\cdot)$ and $\phi(\cdot)$ denote the CDF and PDF of the standard Gaussian random variable, respectively. $\b1{\{ \cdot \} }$ denotes the standard indicator function. For a random variable $Z$, 
$\| Z\|_\infty$ denotes its essential supremum (that is, the infimum of those numbers $z$ such that $P(Z \le z) = 1$.  Boldface letters will denote vectors (e.g., $\by^k=[y_1,\dots,y_k]$) and  continuous-time  process (e.g., $\N=(N_t:t\ge 0)$). We follow the notation of Csisz\'ar and K\"orner \cite{csiszar-korner} for standard information-theoretic quantities. See Karatzas and Shreve~\cite{Karatzas-Shreve} for standard definitions and notations used in stochastic calculus. Unless otherwise stated, all logarithms and exponentiations are base $e$.
\subsection{Definitions}
\label{ssec:defn}
Given a DMC $W \in \cP(\cY|\cX)$, $C$ denotes the capacity of the channel, and 
\begin{equation}
\Pi_W^\ast \eqdef \{ Q \in \cP(\cX) \colon \mI(Q;W) = \mC(W)\}
\end{equation}
denotes the set of capacity-achieving input distributions.
There exists a distribution $q^*$ over $\Y$ such that
for any $P \in \Pi_W^\ast$, 
\begin{equation}
q^*(y) \eqdef \sum_{x \in \cX}P(x)W(y|x).	
\end{equation}
and $q^*$ can be assumed to satisfy $q^*(y)>0$ for all $y\in\Y$~\cite[Corollaries 1 and 2 to Theorem 4.5.1]{Gallager}.\footnote{We assume without loss of generality that $W$ does not contain an all-zero column.}
Define
\begin{gather*}
\ii^*(X,Y)\de\log\frac{W(Y|X)}{q^*(Y)},\nn
\nu_x\de\V[\ii^*(X,Y)|X=x],\nn
\Vm \de \min_{P \in \Pi_W^\ast}\sum_{x\in\X}P(x)\nu_x,\nn
\mV_{\max} \de \max_{P \in \Pi_W^\ast}\sum_{x\in\X}P(x)\nu_x,\nn
\nu_\text{min}\de\min_{x\in\X}\nu_x,\nn
\nu_\text{max}\de\max_{x\in\X}\nu_x,\nn
\ii_{\text{max}}\de \max_{x\in\X,y\in\Y: W(y|x) > 0}|\ii^*(x,y)|
\end{gather*}

Let $\mV_{\min}$ and $\mV_{\max}$ denote $\mV_\varepsilon$ for an arbitrary $\varepsilon \in (0,\frac{1}{2})$ and $\varepsilon \in [\frac{1}{2},1)$, respectively, for notational convenience. 

\begin{definition}
\label{def:arm}
We will call a DMC with\footnote{Note that if $\Vm>0$, then the
capacity of the channel is positive.} $\Vm>0$ \emph{simple-dispersion} if $\Vm=\Vmx$. Otherwise, it is called \emph{compound-dispersion}.
\end{definition}
\begin{remark}
 The set of compound-dispersion DMCs is not empty. As an example, consider\footnote{One can verify that any $p \in [0.8,1)$ satisfies the following.} $p \in (0,1)$ such that 
\begin{equation}
h(p) + (1-p)\log 2 = h(q), 
\label{eq:rem1-0}
\end{equation}
for some $q \in (0,1/2)$, where $h(\cdot)$ denotes the \emph{binary entropy function}, i.e., for any $r \in [0,1]$, $h(r) \eqdef -r \log r - (1-r)\log (1-r)$. Define $\cX \eqdef \{ 0, 1, 2, 3, 4, 5\}$, $\cY \eqdef \{ 0, 1, 2\}$ and $W \in \cP(\cY|\cX)$ as 

\begin{equation}W(y|x) \eqdef  
\left[
\begin{array}{ccc}
p  &  0.5(1-p)  & 0.5(1-p)  \\
0.5(1-p) & p  & 0.5(1-p) \\
0.5(1-p) &  0.5(1-p)  & p \\
q & 1-q & 0 \\
0 & q & 1-q \\
1-q & 0 & q  
\end{array}
\right].
\label{eq:rem1-1}
\end{equation}
One can numerically verify that if $p=0.8$, then $q \approx 0.337$ satisfies \eqref{eq:rem1-0} and the channel defined in \eqref{eq:rem1-1} has $V_{\min} \approx 0.102$, which is attained by the uniform input distribution over the set of input symbols $\{ 3, 4, 5\}$, and $V_{\max} \approx 0.692$, which is attained by the uniform input distribution over the set of input symbols $\{ 0, 1, 2\}$.  Note that for this channel $\nu_\mathrm{min} = V_{\min}$ and $\nu_\mathrm{max} = V_{\max}$. See Strassen~\cite[Sec.~5(ii)]{Strassen} for a similar example.
\eor

\end{remark}

An $(n,R)$ code with ideal feedback for a DMC consists of an encoder $f$, which at the $k$th time instant ($1\le k\le n$) chooses an input $x_k=f(m,y_1\dots,y_{k-1})\in\X$, where $m\in\{1,\dots,\lceil\exp(nR)\rceil\}$ denotes the message to be transmitted, and a decoder $g$, which maps outputs $(y_1,\dots,y_n)$ to  $\hat{m}\in\{1,\dots,\lceil\exp(nR)\rceil\}$. 
Given $\varepsilon \in (0,1)$, define
\begin{equation}
M_{\textnormal{fb}}^\ast(n, \varepsilon) \eqdef \max\left\{\lceil \exp({nR}) \rceil \in \bbR_+ \colon \bar{\mP}_{\me, \textnormal{fb}}(n,R) \leq \varepsilon \right\}, 
\label{eq:Mstar-fb}
\end{equation}
where $\bar{\mP}_{\me}(n,R)$ denotes the minimum average error probability attainable by any $(n,R)$ code with feedback. Similarly, 
\begin{equation}
M^\ast(n, \varepsilon) \eqdef \max\left\{\lceil \exp({nR}) \rceil \in \bbR_+ \colon \bar{\mP}_{\me}(n,R) \leq \varepsilon \right\},   
\label{eq:Mstar}
\end{equation}
where $\bar{\mP}_{\me}(n,R)$ denotes the minimum average error probability attainable by any $(n,R)$ code (without feedback).

\begin{definition}
The \emph{second-order coding rate} of a DMC $W\in\PP(\Y|\X)$ at the average error probability $\varepsilon$ is defined as
\begin{align}
\liminf_{n\to\infty}\frac{\log M^\ast(n, \varepsilon)-nC}{\sqrt{n}}.
\label{EQ:second-order-rate}
\end{align}
\label{def:SOCR}
The second-order coding rate with feedback is defined analogously.
\end{definition}
\subsection{Statement of results}
\label{ssec:results}
Before we state our results, we recall the following result due to Strassen~\cite{Strassen}. For any $W\in\PP(\Y|\X)$ and $\varepsilon \in (0,1)$, Strassen
shows\footnote{Strassen provides a more-refined result, which was 
corrected by Polyanskiy \emph{et al.}~\cite{Polyanskiy}. No correction is 
needed for the weaker result quoted here, however. Strassen states
his result for the maximal error probability criterion then extends
the analysis to the average error probability criterion in Section 5(iii).}
\begin{equation} 
\label{eq:strassen}
\lim_{n\to\infty}\frac{\log M^\ast(n, \varepsilon) -nC}{ \sqrt{n}} = \sqrt{V_{\varepsilon}}\Phi^{-1}(\varepsilon).	
\end{equation}
That is, the second-order coding rate without feedback is 
$\sqrt{V_{\varepsilon}}\Phi^{-1}(\varepsilon)$.
Using timid/bold coding, we shall show that this can be strictly
improved with feedback for any compound-dispersion channel, for
any $0 < \varepsilon < 1$. 

We begin with a preliminary result 
to this effect, which only holds for $0 < \varepsilon < 1/2$
and which does not provide as large of an improvement as
the subsequent result, Theorem~\ref{thrm:ach}.  The advantage
is that its proof does not require any of
the stochastic calculus used in the proofs that follow.

\begin{theorem}[Coarse achievability for compound-dispersion channels]
Fix an arbitrary $\varepsilon \in (0,0.5)$ and consider a 
compound-dispersion channel $W$ with $V_{\min} > 0$. 
Let $\beta = \sqrt{V_{\min}/V_{\max}} < 1$.
Then there exists $1 < \alpha < 1/(2\varepsilon)$ such that
\begin{equation}
\label{eq:alphacondition}
f(\alpha) = \varepsilon (\alpha - 1) - 
               (1 - \beta) \phi(2\sqrt{2} \Phi^{-1}(\alpha \varepsilon))
               \left(\frac{1}{\sqrt{2 \pi}} - \phi(\sqrt{2} \Phi^{-1}(\alpha 
                   \varepsilon))\right) < 0,
\end{equation}
and for any such $\alpha$,
\begin{align}
\liminf_{n\to\infty}\frac{\log M^\ast_{\textnormal{fb}}(n, \varepsilon)-n C}{\sqrt{n }} & \ge \sqrt{V_{\varepsilon}} \Phi^{-1}(\alpha \varepsilon)  \\
      & > \sqrt{V_{\varepsilon}} \Phi^{-1}(\varepsilon).
\end{align}
\label{thrm:ach-bold-timid} 
\end{theorem}
\begin{IEEEproof}
Please see Section~\ref{sec:ach-pf-bold-timid}.
\end{IEEEproof}

The proof proceeds by switching between timid and bold coding
at most once, halfway through the transmission. 
The next result improves upon this by allowing for
a potential switch between timid and bold coding after each time step.

\begin{theorem}[Refined achievability for compound-dispersion channels]
Consider any $W \in \cP(\cY|\cX)$ with $0 < \mV_{\min}$ and let $\beta \eqdef \sqrt{\mV_{\min}/\mV_{\max}}$.
\begin{equation}
\liminf_{n \to \infty}\frac{ \log M^\ast_{\textnormal{fb}}(n, \varepsilon) - n\mC}{\sqrt{n}} \geq 
\begin{cases}
\sqrt{\mV_{\min}}\Phi^{-1}\left( \frac{1}{2 \beta}\varepsilon(1+\beta)\right), &  \varepsilon \in \left(0,\frac{\beta}{1+\beta}\right], \\
\sqrt{\mV_{\max}}\Phi^{-1}\left( \frac{1}{2}[\varepsilon(1+\beta) + (1-\beta)]\right), & \varepsilon \in \left(\frac{\beta}{1+\beta},1\right).
\end{cases}
\label{eq:thrm-ach}
\end{equation}
\label{thrm:ach} 
\end{theorem}
\begin{IEEEproof}
Please see Section~\ref{sec:ach-pf}. 
\end{IEEEproof}

\begin{figure}
\begin{center}
\scalebox{.75}{\includegraphics{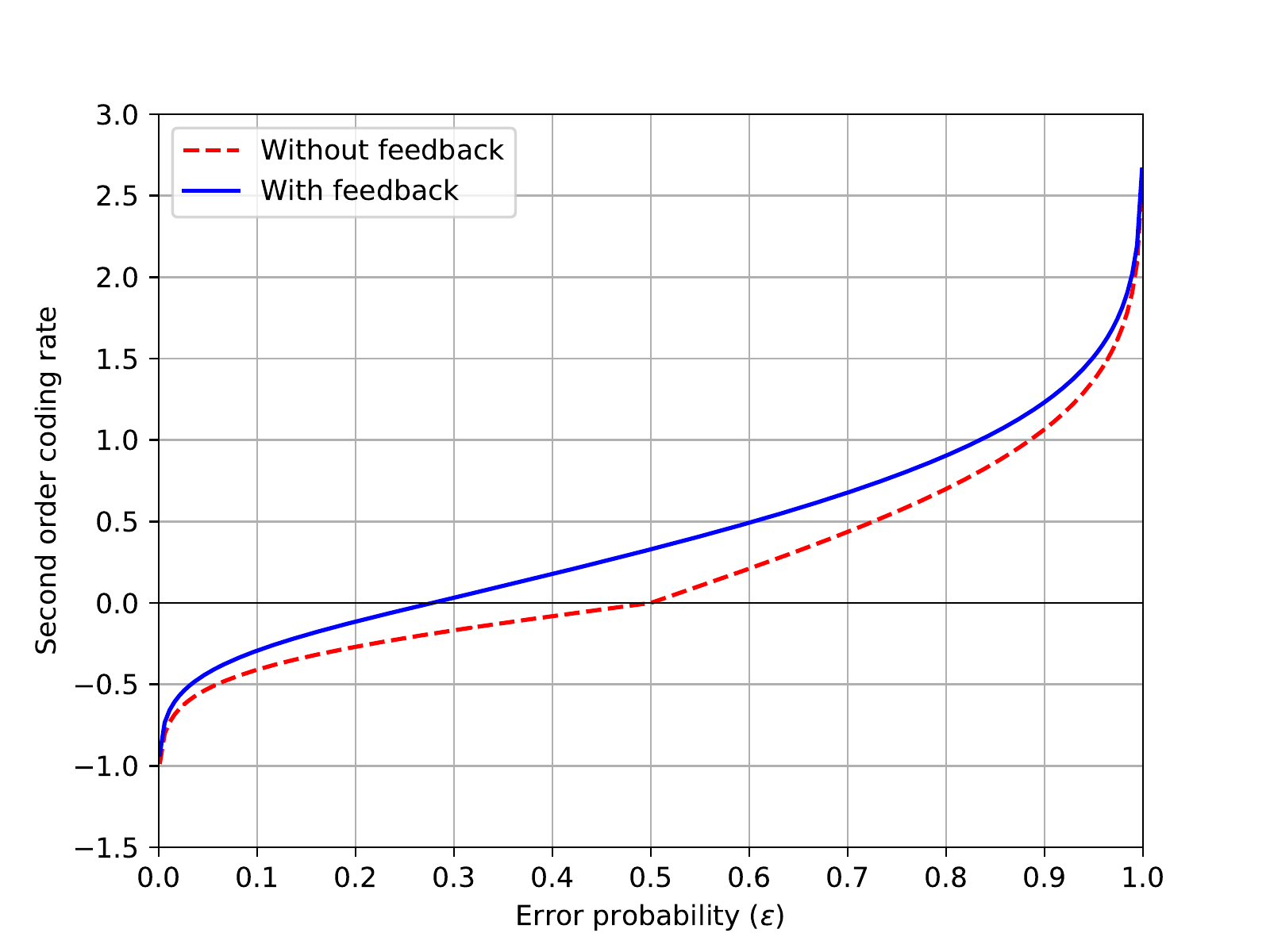}}
\caption{Second order coding rate with and without feedback for
the channel in~(\ref{eq:rem1-1}) with $p = 0.8$. For this channel,
the lower bound in Theorem~\ref{thrm:ach} and the upper bound in 
Theorem~\ref{thm:conv_gen}
coincide, determining the second-order coding rate with feedback.}
\label{fig:improvement}
\end{center}
\end{figure}

Note that the theorem applies to any DMC with $V_{\min} > 0$, but
if $\beta = 1$ (i.e., the channel is simple dispersion),
then (\ref{eq:thrm-ach}) reduces to the
achievability half of~(\ref{eq:strassen}).
The right-hand-side of~(\ref{eq:thrm-ach}) is shown in 
Fig.~\ref{fig:improvement}, 
alongside the second-order coding rate without feedback,
for the channel in (\ref{eq:rem1-1}) with $p = 0.8$ and $q$ selected
to satisfy (\ref{eq:rem1-0}). Note that the range of $\varepsilon$
over which one can approach the capacity from above,
i.e., for which the second-order coding rate is positive,
is enlarged by the presence of feedback.
The right-hand-side of~(\ref{eq:thrm-ach}) is easily verified to
exceed $\sqrt{V_{\varepsilon}}\Phi^{-1}(\varepsilon)$ for
all $\varepsilon$ if the channel is compound-dispersion 
(i.e., $\beta < 1$). The next result shows that if the
channel is not compound-dispersion then feedback
does not improve the second-order coding rate.

\begin{theorem}[Feedback does not improve the second-order coding rate
for simple-dispersion channels]
\label{Thm:con_vmin_vmax}
For any $W\in \PP(\Y|\X)$ with $0<V_\text{min}=V_\text{max}$ (i.e., simple-dispersion) and any $\varepsilon\in(0,1)$,
\begin{equation*}
\limsup_{n\to\infty}\frac{\log M_{\textnormal{fb}}^\ast(n, \varepsilon)-nC}{\sqrt{n}} \le \sqrt{\Vm}\Phi^{-1}\left(\varepsilon \right) = \sqrt{\mV_\varepsilon} \Phi^{-1}\left(\varepsilon \right)
\end{equation*}
\end{theorem}
\begin{IEEEproof}
Please see Section~\ref{sec:proof_con_vmin_vmax}. 
\end{IEEEproof}

The proof of Theorem~\ref{Thm:con_vmin_vmax} 
uses a method of making feedback codes 
``constant-composition,'' which is inspired by Fong and Tan's
work on parallel Gaussian channels~\cite{Fong-Tan}.
Fong and Tan have also noted that their techniques can
be applied to DMCs to obtain something like 
Theorem~\ref{Thm:con_vmin_vmax}~\cite{Fong-Tan-PC}.

If the channel is compound dispersion, then feedback improves
the second-order coding rate, and Theorem~\ref{thrm:ach} (along with
(\ref{eq:strassen})) provides
a lower bound on the size of the improvement. The next theorem
provides a comparable upper bound.

\begin{theorem}[Impossibility for compound-dispersion channels]
\label{thm:conv_gen}
Consider any $W \in \cP(\cY|\cX)$ with $0 < \nmn$ and let $\lambda \eqdef \sqrt{\nmn/\nmx}$. Then
\begin{equation}
\limsup_{n \to \infty}\frac{\log M^\ast_{\textnormal{fb}}(n, \varepsilon) - n\mC}{\sqrt{n}} \leq 
\begin{cases}
\sqrt{\nmn}\Phi^{-1}\left( \frac{1}{2 \lambda}\varepsilon(1+\lambda)\right), &  \varepsilon \in \left(0,\frac{\lambda}{1+\lambda}\right], \\
\sqrt{\nmx}\Phi^{-1}\left( \frac{1}{2}[\varepsilon(1+\lambda) + (1-\lambda)]\right), & \varepsilon \in \left(\frac{\lambda}{1+\lambda},1\right).
\end{cases}
\end{equation}
\end{theorem}
\begin{IEEEproof}
Please see Section~\ref{sec:proof_conv_gen}. 
\end{IEEEproof}

The upper bound in Theorem~\ref{thm:conv_gen} equals the achievability result in Theorem~\ref{thrm:ach} but with $\nmn$ and $\nmx$ replacing $\Vm$ and $\Vmx$, respectively. Thus the two results are similar in spirit. Both, in fact, use McNamara's scheme in~(\ref{eq:bangbang}). However, the range of values that the diffusion coefficient can assume is larger for the upper bound ($[\sqrt{\nmn}, \sqrt{\nmx}]$) than for the lower bound ($[\sqrt{\Vm}, \sqrt{\Vmx}]$). 
For the channel in~(\ref{eq:rem1-1}), $\nmx = \Vmx$ and $\nmn = \Vm$,
so the upper and lower bound coincide and the second-order coding
rate with feedback is determined (and is depicted in 
Fig.~\ref{fig:improvement}). 
The two bounds do not coincide in general, however.

Finally, we consider very noisy channels (VNCs). For our purposes,
a very noisy channel is one of the form
\begin{align}
W_\zeta(y|x)=\Gamma(y)\left(1+\zeta\lambda(x,y)\right),
\label{EQ:VNC}
\end{align}
where $\Gamma$ is a probability distribution on the output alphabet  
$\Y$ such that $\Gamma(y)>0$ for all $y$, $\lambda(x,y)$ satisfies 
\begin{align}
\sum_{y\in\Y}\Gamma(y)\lambda(x,y)=0
\label{EQ:Lambda_Sum}
\end{align} 
for all $x\in\X$, and $\zeta$ is infinitesimally small.
In the very noisy limit, i.e., as $\zeta$ tends to zero, 
$\mV_{\min}$ and $\Vmx$ converge together and the
channel behaves as one with simple dispersion. In light
of Theorem~\ref{Thm:con_vmin_vmax}, one therefore expects feedback not
to improve the second-order coding rate in the very noisy
limit. Since $\mV_{\min}$ 
and $\Vmx$ are only equal in the limit (when suitably
scaled), the result does not follow from 
Theorem~\ref{Thm:con_vmin_vmax}, however.
Since $\sqrt{\nmn}$ and $\sqrt{\nmx}$ do not necessarily 
converge together, the result does not follow from
Theorem~\ref{thm:conv_gen} either.

\begin{theorem}[Feedback does not improve the second-order coding rate
in the very noisy limit]
\label{Thm:conv_vnc}
Consider a channel family $W_\zeta\in \cP(\cY|\cX)$ of the form $W_\zeta(y|x)=\Gamma(y)\left(1+\zeta\lambda(x,y)\right),$ with $\Gamma\in\cP(\cY)$. Let $C_\zeta$, $V_{\text{min},\zeta}$, $V_{\text{max},\zeta}$, and $\log M^\ast_{\textnormal{fb},\zeta}(n, \varepsilon)$  denote $C$, $\Vm$, $\Vmx$, and $M^\ast_{\textnormal{fb}}(n, \varepsilon)$, respectively, for the channel $W_\zeta\in \cP(\cY|\cX)$. If 
$$
\max_{P\in\mathcal{P}(\X)}\frac{1}{2}\sum_{y\in\Y}\Gamma(y)\Bigg(\left.\sum_{x\in \X}P(x)\lambda^2(x,y)-\left( \sum_{x\in\X}P(x)\lambda(x,y)\right)^2 \right) > 0,
$$
which ensures that $C_\zeta > 0$ for all sufficiently small $\zeta$, then
\begin{align*}
\limsup_{\zeta\to 0}\limsup_{n\to\infty}\frac{\log M^\ast_{\textnormal{fb},\zeta}(n, \varepsilon) -nC_\zeta}{\sqrt{nV_{\text{min},\zeta}}}\le \Phi^{-1}(\varepsilon), &\quad \varepsilon \in \left(0,\frac{1}{2}\right]\\
\limsup_{\zeta\to 0}\limsup_{n\to\infty}\frac{\log M^\ast_{\textnormal{fb},\zeta}(n, \varepsilon) -nC_\zeta}{\sqrt{nV_{\text{max},\zeta}}}\le \Phi^{-1}(\varepsilon), &\quad \varepsilon \in \left(\frac{1}{2},1\right).
\end{align*}
\end{theorem}
\begin{IEEEproof}
Please see Section~\ref{sec:vnc}. 
\end{IEEEproof}
One can also show that feedback does not improve the high-rate
error exponent or moderate deviations performance of 
VNCs~\cite{Shende_VNC}.
Note that very noisy channels are unusual in that their reliability
function is known at all rates~\cite{Gallager, LeeWinick:DMC}.

The next five sections contain the proofs of 
Theorems~\ref{thrm:ach-bold-timid}
through~\ref{Thm:conv_vnc}, respectively.

\section{Proof of Theorem~\ref{thrm:ach-bold-timid}}
\label{sec:ach-pf-bold-timid}

Note that $f(\cdot)$ is continuous on $[1,\infty)$ and
$f(1) < 0$. Hence there exists $1 < \alpha < 1/(2\varepsilon)$
with $f(\alpha) < 0$ and we fix any such $\alpha$ in what
follows. Define
\begin{equation}
\nu = \sqrt{2} \Phi^{-1}(\alpha \varepsilon) < 0.
\end{equation}

We shall use Lemma~\ref{lemma:shannon:achievability:FB} in the Appendix.
Note that we only require that
(\ref{eq:achievability:control:hypothesis}) holds with the limit 
superior taken along the even
integers. Accordingly, suppose that $n$ is even. Let $Q_{\max}$
denote a distribution on $\PP(\mathcal{X})$ that attains
$V_{\max}$, and define $Q_{\min}$ similarly. Select the controller
$F$ as follows
\begin{equation}
F(x^k,y^k) = \begin{cases}
Q_{\min} & \text{if $k \le n/2$} \\
Q_{\min} & \text{if $k > n/2$ and 
                 $\log \frac{W(y^{n/2}|x^{n/2})}{q^*(y^{n/2})} >
                       \frac{nC}{2} + \nu \sqrt{\frac{n\Vm }{2}}$} \\
Q_{\max} & \text{if $k > n/2$ and 
                 $\log \frac{W(y^{n/2}|x^{n/2})}{q^*(y^{n/2})} \le 
                       \frac{nC}{2} + \nu \sqrt{\frac{n\Vm}{2}}$}.
\end{cases}
\end{equation}
Note that $FW = q^* \times q^* \times \cdots q^*$.
For convenience we define
\begin{equation}
\Gamma_n = (F \circ W) \left(\sum_{k = 1}^n \log \frac{W(Y_k|X_k)}
                         {q^*(Y_k)} \le nC + \sqrt{n \Vm} \Phi^{-1}(\alpha
                             \varepsilon)\right).
\end{equation}

Let $\underline{G}_n$ denote the CDF of $\frac{1}{\sqrt{(n/2) V_{\min}}}\sum_{i=1}^{n/2}\left[ \log \frac{W(Y_i|X_i)}{q^\ast(Y_i)} - C\right]$ when $\left\{ \log \frac{W(Y_i|X_i)}{q^\ast(Y_i)} \right\}_{i=1}^{n/2}$ are i.i.d.\ with distribution $Q_{\min}\circ W$. 
Similarly, let $\overline{G}_n$ denote the distribution of $\frac{1}{\sqrt{(n/2) V_{\min}}}\sum_{i=1}^{n/2}\left[ \log \frac{W(Y_i|X_i)}{q^\ast(Y_i)} - C \right]$ when $\left\{ \log \frac{W(Y_i|X_i)}{q^\ast(Y_i)} \right\}_{i=1}^{n/2}$ are i.i.d.\ with distribution $Q_{\max} \circ W$. 
We have
\begin{align}
\Gamma_n & = \int\limits_{\nu}^\infty \underline{G}_n\left( \nu - x \right)d\underline{G}_n(x) + \int\limits_{-\infty}^{\nu} \overline{G}_n\left( \nu- x \right)d\underline{G}_n(x) \nonumber \\
& = \underline{G}_{2n}\left( \Phi^{-1}(\alpha \varepsilon)\right) - \int\limits_{-\infty}^{\nu}\left[ \underline{G}_n\left(\nu - x \right) - \overline{G}_n\left(\nu- x \right)\right]d\underline{G}_n(x). \label{eq:Pe1-final1}
\end{align}
From the Berry-Esseen theorem\footnote{For the sake of notational convenience, we take the universal constant in the theorem as $1/2$, although this is not the best known constant for the case of i.i.d. random variables. See \cite{korolev2010} for a survey of the best known constants in the Berry-Esseen theorem.} \cite{berry41,esseen45}, along with a first-order Taylor series approximation, we deduce that 
\begin{equation}
\underline{G}_{2n}\left( \Phi^{-1}(\alpha \varepsilon) \right) \leq \alpha \varepsilon + \frac{\underline{\kappa}}{2\sqrt{n}}, \label{eq:Pe1-CLT-1}
\end{equation}
where $\underline{\kappa} \eqdef  \mE_{Q_{\min} \circ W}\left[ \left| \log W(Y|X)/q^\ast(Y) - C\right|^3 \right]/V_{\min}^{3/2} + 1$. Another application of the Berry-Esseen theorem implies that for any $x \in \bbR$, 
\begin{align}
\left| \underline{G}_{n}\left( \nu - x \right) - \Phi\left( \nu - x \right)\right| \le \frac{\underline{\kappa}}{\sqrt{2n}}, \label{eq:Pe1-CLT-2}\\
\left|\overline{G}_{n}\left( \nu - x \right) - \Phi\left( \beta\left[\nu - x \right] \right)\right|  \le  \frac{\overline{\kappa}}{\sqrt{2n}}, \label{eq:Pe1-CLT-3}
\end{align}
where $\overline{\kappa} \eqdef  \mE_{Q_{\max} \circ W}\left[ \left| \log W(Y|X)/q^\ast(Y) - C \right|^3 \right]/V_{\max}^{3/2} + 1$. Equations \eqref{eq:Pe1-CLT-2} and \eqref{eq:Pe1-CLT-3} imply that 
\begin{align}
\int\limits_{-\infty}^{\nu}\left[\underline{G}_n(\nu -x) - \overline{G}_n(\nu -x)\right] d\underline{G}_n(x) 
& \geq \int\limits_{-\infty}^{\nu}\left[ \Phi(\nu - x) - \Phi\left( \beta \left[ \nu - x \right]\right)\right]d\underline{G}_n(x)  - \frac{\underline{\kappa} + \overline{\kappa}}{\sqrt{2n}} \\
& =  \int\limits_{-\infty}^{\nu}\underline{G}_n(x)\left[ \phi(\nu - x) - \beta\phi\left( \beta \left[ \nu - x \right]\right)\right]dx - \frac{\underline{\kappa} + \overline{\kappa}}{\sqrt{2n}}\label{eq:Pe1-CLT-4} \\
& \geq \int\limits_{-\infty}^{\nu}\Phi(x)\left[ \phi(\nu - x) - \beta\phi\left( \beta \left[ \nu - x \right]\right)\right]dx - \frac{3\underline{\kappa} + \overline{\kappa}}{\sqrt{2n}} \label{eq:Pe1-CLT-5}\\
& = \int\limits_{-\infty}^{\nu}\phi(x)\left[ \Phi(\nu - x) - \Phi(\beta [\nu-x])\right]dx - \frac{3\underline{\kappa} + \overline{\kappa}}{\sqrt{2n}}, \label{eq:Pe1-CLT-6}
\end{align}
where \eqref{eq:Pe1-CLT-4} and \eqref{eq:Pe1-CLT-6} follow from integration by parts and \eqref{eq:Pe1-CLT-5} follows from the Berry-Esseen theorem. 
We continue as follows
\begin{align}
\int\limits_{-\infty}^{\nu}\phi(x)\left[ \Phi(\nu - x) - \Phi(\beta [\nu-x])\right]dx & = \int\limits_{0}^{\infty}\phi(\nu-z)\int\limits_{\beta z}^z \phi(\zeta) d\zeta dz \label{eq:Pe1-CLT-7.5}\\
& \geq (1-\beta)\int\limits_{0}^{\infty}\phi(\nu - z)z \phi(z)dz \\
& \geq (1-\beta)\phi(2\nu)\int\limits_{0}^{-\nu}z\phi(z)dz \\
& = (1-\beta)\phi(2\nu)\left( \frac{1}{\sqrt{2 \pi}} - \phi(\nu)\right) \label{eq:Pe1-CLT-8}.
\end{align}
By plugging \eqref{eq:Pe1-CLT-8} into \eqref{eq:Pe1-CLT-6}, and recalling \eqref{eq:Pe1-final1} and \eqref{eq:Pe1-CLT-1}, we deduce that 
\begin{equation}
\Gamma_n \le f(\alpha) + \varepsilon + \frac{4\underline{\kappa} + \overline{\kappa}}{ \sqrt{2n}}.
\end{equation}
Thus for all sufficiently large (and even) $n$, we have
\begin{equation}
\Gamma_n < \varepsilon.
\end{equation}
So by Lemma~\ref{lemma:shannon:achievability:FB}, 
\begin{equation}
\liminf_{n \rightarrow \infty} \frac{\log M^\ast_{\textnormal{fb}}(n,\varepsilon) - n C}{\sqrt{n}} \ge \sqrt{V_{\min}} \Phi^{-1}(\alpha \varepsilon).
\end{equation}
\Qed

\begin{remark}
Although Theorem~\ref{thrm:ach-bold-timid} uses feedback only at
a single epoch, it still provides a strict improvement over the
best non-feedback code. It is possible to prove a version of 
Theorem~\ref{thrm:ach-bold-timid} for large $\varepsilon$, but we shall 
not pursue this here because our aim with Theorem~\ref{thrm:ach-bold-timid}
is only to elucidate the idea behind timid/bold coding while
avoiding the diffusion machinery used in our main achievability
result, Theorem~\ref{thrm:ach}. Theorem~\ref{thrm:ach} takes timid/bold
coding to its natural limit by allowing the encoder to switch between
timid and bold signaling schemes after each time-step. 
\label{rem:bold-timid}
\eor
\end{remark}

\section{Proof of Theorem~\ref{thrm:ach}}
\label{sec:ach-pf}

Following {\O}ksendal (e.g., \cite[Def.~7.1.1]{oksendal00}), we define a one-dimensional, time-homogeneous \emph{It\^o diffusion} as follows.
\begin{definition}[It\^o diffusion] A time-homogeneous It\^o diffusion is a stochastic process $\mathbf{X}$ satisfying a stochastic differential equation of the form 
\begin{equation}
X_t = x_0 + \int\limits_{0}^t b(X_s)\,ds + \int\limits_{0}^t \sigma(X_s)\,dB_s,
\label{eq:Ito-1}
\end{equation}
for some one-dimensional Brownian Motion $\B$ defined on the same sample space, where $b \colon \bbR \to \bbR$ and $\sigma \colon \bbR \to \bbR$ are measurable functions that satisfy 
\begin{equation}
|b(x)-b(y)| + |\sigma(x) - \sigma(y)| \leq D |x-y|, \, \forall \, x,y \in \bbR,
\label{eq:Ito-2}
\end{equation}
for some constant $D \in \bbR^+$.
\label{def:Ito-diff}
\end{definition}
\begin{remark}
Since \eqref{eq:Ito-2} ensures that the conditions in \cite[Theorem~5.2.1]{oksendal00} are satisfied, \eqref{eq:Ito-1} has a unique solution.
\end{remark}
\subsection{A convergence result}
Let $\{ Z_{i,k}\}_{k=1}^{\infty}$, $i \in \{ 0,1\}$ denote i.i.d.\ sequences of bounded random variables, which are also independent of each other, such that for any $k \in \bbZ^+$, 
\begin{align}
\E[Z_{0,k}]=\mE[Z_{1,k}] & =0, \label{eq:markov-moments-1}\\
\E[Z_{1,k}^2] & =1, \label{eq:markov-moments-2}\\
\E[Z_{0,k}^2] & =\beta^2, \label{eq:markov-moments-3}	
\end{align}
with $\beta \in (0,1)$. Given any $\delta \in (0,1]$ and $x \in [0,\delta]$,  
define
\begin{equation}
\alpha_{\delta}(x) \eqdef \frac{1}{1-\beta^2}\left( \left[ 1 - x \left(\frac{1-\beta}{\delta}\right)\right]^2 - \beta^2 \right).\label{eq:markov-alpha}
\end{equation}
Via direct computation, one can verify that 
\begin{equation}
\alpha_{\delta}(x) \in [0,1],	
\end{equation}
for the given range of $\delta$ and $x$. Let $\mu_i$ denote the law of $Z_{i,1}$ for $i \in \{ 0,1\}$. Define the probability measure 
\begin{equation}
\mu_{\delta, x} \eqdef (1-\alpha_{\delta}(x))\mu_0 + \alpha_{\delta}(x)\mu_1.	
\end{equation}
For any $\varepsilon \in (0,1)$,  define
\begin{equation}
s(\varepsilon) \eqdef 
\begin{cases}
-\beta \Phi^{-1}\left( \frac{1}{2 \beta}\varepsilon (1+\beta)\right), &\varepsilon \in (0, \frac{\beta}{1+\beta}], \\ 
-\Phi^{-1}\left( \frac{1}{2}[\varepsilon (1+\beta)+(1-\beta)]\right), &\varepsilon \in (\frac{\beta}{1+\beta}, 1).
\end{cases}
\label{eq:markov-s0}
\end{equation}
For any $\varepsilon \in (0,1)$ and $n \in \bbZ^+$, 
\begin{align}
S_{0}^{\delta, \varepsilon,n} & \eqdef s(\varepsilon)\sqrt{n}, \label{eq:markov-S_nk-1}\\
S_{k+1}^{\delta, \varepsilon,n} & \eqdef S_{k}^{\delta, \varepsilon,n} + \b1\left\{S_{k}^{\delta, \varepsilon,n} \leq 0 \right\}Z_{1,k+1} + \b1\left\{S_{k}^{\delta, \varepsilon,n} > \delta \sqrt{n}\right\}Z_{0,k+1} + \b1\left\{ 0< S_{k}^{\delta, \varepsilon,n} \leq \delta \sqrt{n}\right\}Z_{2,k+1}, \label{eq:markov-S_nk-2}
\end{align}
for all $k \in \bbZ^+$, where $Z_{2,k+1}$ has distribution
$\mu_{\delta, S_{k}^{\delta, \varepsilon,n}/\sqrt{n}}$ and is
independent of $\{ Z_{i,j}\}_{j=1}^\infty$, $i \in \{ 0,1\}$ and
$\{ Z_{2,j}\}_{j=1}^{k}.$
\begin{proposition}
Consider any $\varepsilon \in (0,1)$. For any $\kappa \in \bbR^+$, there exist $\delta_\mo \in (0,1)$ and $n_\mo \in \bbZ^+$ such that for all $n \geq n_\mo$, 
\begin{equation}
\Pr\left(\frac{1}{\sqrt{n}}S_{n}^{\delta_\mo, \varepsilon,n}\leq 0\right) \leq \varepsilon + \kappa.	
\end{equation}
\label{prop:conv}
\end{proposition}
\vspace{-0.5cm}
\begin{IEEEproof}
Similar to \cite[p.~43]{Kushner}, we interpolate the discrete-time Markov process defined in \eqref{eq:markov-S_nk-1} and \eqref{eq:markov-S_nk-2} as follows
\begin{equation}
\xi^{\varepsilon, \delta,n}_{t} \eqdef \frac{1}{\sqrt{n}}S^{\delta, \varepsilon,n}_{[nt]}, \label{eq:markov-interpol}
\end{equation}
for any $t \in \bbR_+$, where $[nt]$ denotes the integer part of $nt$. We prove the claim by investigating the limiting behavior of $\xi^{\varepsilon,\delta,n}_t$ as $\delta \to 0$ and $n \to \infty$. To this end, we use several stochastic processes, which are defined next. 

For any $\delta \in (0,1]$, define $\sigma_\delta : \bbR \to \bbR$ as
\begin{equation}
\sigma_\delta (x) \eqdef 
\begin{cases}
1, & x \leq 0, \\
1 - x \left( \frac{1-\beta}{\delta}\right), &  0 \leq x \leq \delta, \\
\beta, & x \geq \delta.  
\end{cases}
\label{eq:markov-sigma-delta}
\end{equation}
Clearly, $\sigma_{\delta}(\cdot)$ is Lipschitz continuous, positive and bounded. For any $\varepsilon \in (0,1)$, we use \eqref{eq:markov-sigma-delta} to define an It\^o diffusion $\xi^{\varepsilon, \delta}_t$ that is the solution of the following stochastic differential equation:
\begin{equation}
\xi^{\varepsilon,\delta}_t = \xi^{\varepsilon,\delta}_0 + \int\limits_{0}^t \sigma_\delta(\xi^{\varepsilon,\delta}_s)\,dB_s,\label{eq:markov-xi-delta-1}
\end{equation}
with $\xi^{\varepsilon,\delta}_0 \eqdef s(\varepsilon)$. Further, define $\bar{\sigma}:\bbR \to \bbR$ with 
\begin{equation}
\bar{\sigma}(x) \eqdef \b1\{ x \leq 0 \} + \beta \b1\{ x > 0\},	
\end{equation}
and let $\xi^{\varepsilon,0}_t$ be the solution of the following stochastic differential equation
\begin{equation}
\xi^{\varepsilon,0}_t = \xi^{\varepsilon,0}_0 + \int\limits_{0}^t \bar{\sigma}(\xi^{\varepsilon,0}_s)\,dB_s,\label{eq:markov-xi-delta-0}
\end{equation}
with $\xi^{\varepsilon,0}_0 \eqdef s(\varepsilon)$. Existence of a (weak) solution of \eqref{eq:markov-xi-delta-0} can be verified by using~\cite[Theorem~23.1]{Kallenberg}. Further, an expression for the transition probabilities of the Markov process $\xi^{\varepsilon,0}_t$, denoted by $P_t(x,y)$, is known \cite{kulinich67},
\begin{equation}
P_t(x,y) =  \frac{1}{\sqrt{2\pi t}}
\begin{cases}
\frac{1}{\beta}\exp\left(-(x-y)^2/2\beta^2 t\right)-\frac{(\beta-1)}{\beta(\beta+1)}\exp\left({-(x+y)^2/2 \beta^2 t}\right), & (x,y) \in \bbR^+ \times \bbR^+, \\
\frac{2\beta}{(\beta+1)} \exp\left({-(x-\beta y)^2/2 \beta^2 t}\right), & (x,y) \in \bbR^+ \times \bbR^-, \\
\frac{2 }{\beta(\beta+1)}\exp\left({-(\beta x -y)^2/2 \beta^2 t}\right), & (x,y) \in \bbR^- \times \bbR^+, \\
\exp\left({-(x-y)^2/2t}\right)+\frac{(\beta-1)}{(\beta+1)}\exp\left({-(x+y)^2/2t}\right), & (x,y) \in \bbR^- \times \bbR^-.
\end{cases}
\label{eq:markov-xi-delta-0-trans}
\end{equation}

In Lemmas~\ref{lem:markov-lem-1} and \ref{lem:markov-lem-2} to follow, the mode of convergence is the weak convergence of probability measures in the space of right-continuous functions with left limits defined on $[0, 1]$, i.e., $D[0,1]$, endowed with the Skorohod topology (e.g., \cite[Section~12]{billingsley99}). 

\begin{Lemma}
\begin{equation}
\boldsymbol{\xi}^{\varepsilon,\delta} \xrightarrow{w.}	\boldsymbol{\xi}^{\varepsilon,0}, \textnormal{ as } \delta \to 0.
\end{equation}
\label{lem:markov-lem-1}
\end{Lemma}
\vspace{-0.75cm}
\begin{IEEEproof}
The claim follows from a convergence result due to Kulinich~\cite[Theorem~2]{kulinich83}. To verify the conditions of this theorem for our case, we note that the function $f_\delta$ in~\cite[p.~856]{kulinich83} can be taken to be $f_\delta(x)=x$, either by direct calculation or by noticing the fact that the It\^o diffusion $\xi^{\varepsilon,\delta}_t$ is in its natural scale. The condition regarding $f^\prime_\delta(\cdot) \sigma_\delta(\cdot)$ is satisfied, since 
\begin{equation}
\beta \leq f^\prime_\delta(x) \sigma_\delta(x) \leq 1,	
\end{equation}
for all $\delta \in (0,1]$ and $x \in \bbR$. Further, the condition
\begin{equation}
\lim_{K \to \infty}\lim_{\delta \to 0} \Pr(|f_\delta(\xi^{\varepsilon,\delta}_0)| > K) = 0,
\end{equation}
is also clearly satisfied since 
\begin{equation}
f_\delta(\xi^{\varepsilon,\delta}_0)=s(\varepsilon) \in \bbR.	
\end{equation}
Finally, the condition regarding the function $G_\delta$, which is defined in~\cite[p.~857]{kulinich83}, can be verified to hold for our case, since for any $x \in \bbR$, we have 
\begin{align}
\lim_{\delta \to 0} G_\delta(x) & = \lim_{\delta \to 0}\int\limits_{0}^x\frac{\,du}{\sigma^2_\delta(u)} \\
& = \frac{x}{\bar{\sigma}^2(x)},
\end{align}
via direct calculation. Hence, we can apply~\cite[Theorem~2]{kulinich83} to deduce the assertion, since the generalized diffusion used in this theorem, which is defined in~\cite[Eq.~(3)]{kulinich83}, reduces to $\xi^{0}_t$ in our case.
\end{IEEEproof}

\begin{Lemma}
For any $\delta \in (0,1]$, 
\begin{align}
\boldsymbol{\xi}^{\varepsilon,\delta,n} \xrightarrow{w.}\boldsymbol{\xi}^{\varepsilon,\delta} \textnormal{ as }n \to \infty. 	
\end{align}
\label{lem:markov-lem-2}
\end{Lemma}
\vspace{-0.75cm}
\begin{IEEEproof}
The claim follows from a convergence result of Kushner~\cite[Theorem~1]{Kushner}. Specifically, we apply this theorem with the Markov chain
\begin{equation}
\left\{ \frac{1}{\sqrt{n}}S_{k}^{\delta, \varepsilon,n}\right\}_{k=0}^\infty,	
\end{equation}
$\mathcal{F}_{k,n}$ denoting the sigma-algebra generated by $\frac{S_{i}^{\delta, \varepsilon,n}}{\sqrt{n}}$ for all $i \leq k$, and the sequence of positive real numbers $\delta t_{i}^n = \frac{1}{n}$. The definition of $S_{k}^{\delta, \varepsilon,n}$, along with \eqref{eq:markov-sigma-delta} and elementary algebra, ensures that for any $n \in \bbZ^+$, we have
\begin{equation}
\E \left[ \left( S_{k+1}^{\delta, \varepsilon,n}-S_{k}^{\delta, \varepsilon,n}\right)^2 \bigg| \mathcal{F}_{k,n}\right] =\sigma^2_{\delta}\left(  \frac{S_{k}^{\delta, \varepsilon, n}}{\sqrt{n}} \right) \textnormal{ (a.s.)},
\end{equation}
for all $t \in \bbR_+$ and $k \in \{ 0, \ldots, [nt]\}$, and hence the condition in \cite[Eq.~(1)]{Kushner} is satisfied. The proof will be complete if we can verify that the six assumptions of Kushner~\cite[pg.~42]{Kushner} are satisfied for  our case. Indeed, except (A4) and (A6), these assumptions trivially hold with the aforementioned choices. (A6) is evidently true since $\xi^{\varepsilon,\delta}_t$ is the unique (strong) solution of \eqref{eq:markov-xi-delta-1}, whereas (A6) only requires \eqref{eq:markov-xi-delta-1} to possess a unique weak solution (e.g., \cite[Chapter~5.3]{oksendal00}). To verify (A4), let $K \in \bbR^+$ be a constant such that 
\begin{equation}
\max\{ |Z_{0,1}|, |Z_{1,1}| \} \leq K \textnormal{ (a.s.)},	
\end{equation}
whose existence is ensured by the boundedness of the random variables. From the definition of $S_{k}^{\delta, \varepsilon,n}$, one can verify that for any $t \in \bbR^+$,
\begin{align}
0 & \leq \E\left[ \sum_{k=0}^{[nt]} \left|  \frac{S_{k+1}^{\delta, \varepsilon,n}-S_{k}^{\delta, \varepsilon,n}}{\sqrt{n}} \right|^{3}\right] \\
& \leq \frac{1}{n^{3/2}}K^3([nt]+1) \to 0, \textnormal{ as } n \to \infty. 
\label{eq:markov-A6}
\end{align}
Evidently, \eqref{eq:markov-A6} implies (A4) and hence we can apply \cite[Theorem~1]{Kushner} to infer the assertion.
\end{IEEEproof}
In order to conclude the proof, it suffices to note that 
\begin{align}
\lim_{\delta \to 0}\Pr(\xi^{\varepsilon,\delta}_1 \leq 0) &=  \Pr(\xi^{\varepsilon, 0}_1 \leq 0), \label{eq:markov-final-1}\\
 \lim_{n \to \infty} \Pr(\xi^{\varepsilon,\delta,n}_1 \leq 0) &= \Pr(\xi^{\varepsilon,\delta}_1 \leq 0),  \, \forall \, \delta \in (0,1], \label{eq:markov-final-2}\\
\Pr(\xi^{\varepsilon,0}_1 \leq 0)  &= \varepsilon,  \label{eq:markov-final-3}
\end{align}
where \eqref{eq:markov-final-1} and \eqref{eq:markov-final-2} follow from Lemmas~\ref{lem:markov-lem-1} and \ref{lem:markov-lem-2}, respectively, along with \cite[Theorem~12.5]{billingsley99}, whereas \eqref{eq:markov-final-3} follows from an elementary calculation by using \eqref{eq:markov-xi-delta-0-trans}.
\end{IEEEproof}
\subsection{Proof of Theorem~\ref{thrm:ach}}
Fix any $\varepsilon \in (0, 1)$. If $\beta = 1$ then the
result is implied by~(\ref{eq:strassen}). Otherwise, assume that
\begin{equation}
\beta = \sqrt{\frac{\mV_{\min}}{\mV_{\max}}} \in (0,1).	
\end{equation}
Choose some $0< \kappa < \frac{\varepsilon}{2}$ that also satisfies 
\begin{equation}
\kappa \leq \frac{\left[\varepsilon- \frac{\beta}{1+\beta}\right]}{4}
\end{equation}
if $\varepsilon > \frac{\beta}{1+\beta}$. 
Define $r : (0,1) \mapsto \mathbb{R}$ as
\begin{equation}
r(a) \eqdef 
\begin{cases}
\sqrt{\mV_{\min}}\Phi^{-1}\left( \frac{a(1+\beta)}{2\beta}\right), & 0< a \leq \frac{\beta}{1+\beta}, \\
\sqrt{\mV_{\max}}\Phi^{-1}\left( \frac{a(1+\beta)+(1-\beta)}{2}\right), & \frac{\beta}{1+\beta} < a < 1.
\end{cases}
\label{eq:rn}
\end{equation}
Using $r(\cdot)$, define 
\begin{equation}
R_n(\cdot) \eqdef \mC + \frac{r(\cdot)}{\sqrt{n}}.
\label{eq:Rn}
\end{equation}
Again we shall use 
Lemma~\ref{lemma:shannon:achievability:FB} in the Appendix. 
To this end, define
the controller $F_\ell$ via
\begin{equation}
F_\ell(x^{k-1},y^{k-1}) = \begin{cases}
Q_{\max}, & \sum\limits_{j=1}^{k-1}\left[ \log \frac{W(y_j|x_j)}{q^\ast(y_j)} - \mC \right] \leq \sqrt{n} r(\varepsilon-\kappa), \\
Q_{\min}, &  \sum\limits_{j=1}^{k-1}\left[ \log \frac{W(y_j|x_j)}{q^\ast(y_j)} - \mC \right]  > \sqrt{n} r(\varepsilon-\kappa) + \frac{1}{\ell}\sqrt{nV_{\max}}, \\
\alpha_{\ell,k}Q_{\max}+(1-\alpha_{\ell,k})Q_{\min}, & \textnormal{else},
\end{cases}
\label{eq:Qxiyi}
\end{equation}
where 
\begin{equation}
\alpha_{\ell,k} \eqdef \alpha_{1/\ell}\left( -\frac{r(\varepsilon-\kappa)}{\sqrt{\mV_{\max}}} + \frac{1}{\sqrt{n \mV_{\max}}}\sum_{j=1}^{k-1} \left[ \log \frac{W(y_j|x_j)}{q^\ast(y_j)} - \mC \right] \right),
\end{equation}
by using the function defined in \eqref{eq:markov-alpha} and with a slight abuse of notation, we let
\begin{equation}
\sum_{j=1}^{0}\left[ \log \frac{W(y_j|x_j)}{q^\ast(y_j)} - \mC \right]=0. 	
\end{equation}

By Proposition~\ref{prop:conv}, there exists $\ell_0$ in $\mathbb{Z}^+$ and
$n_0$ in $\mathbb{Z}^+$ such that if $n \ge n_0$ and $\ell \ge \ell_0$,
\begin{equation}
(F_\ell \circ W)\left( \frac{1}{\sqrt{n}} \sum_{k = 1}^n \left(
       \log \frac{W(Y_k|X_k)}{q^*(Y_k)} - C\right) - r(\varepsilon - 
        \kappa) \le 0 \right) \le \varepsilon - \frac{\kappa}{2}.
\end{equation}
Lemma~\ref{lemma:shannon:achievability:FB} then implies that
\begin{equation}
\liminf_{n \rightarrow \infty} \frac{\log M_{\textnormal{fb}}^\ast(n, \varepsilon) - n \mC}{\sqrt{n}} \ge r(\varepsilon-\kappa). \label{eq:final} 
\end{equation}
Since $r(\cdot)$ is continuous and $\kappa >0$ is arbitrary, the result follows.\hfill\IEEEQED 

\section{Proof of  Theorem~\ref{Thm:con_vmin_vmax}}
\label{sec:proof_con_vmin_vmax}

Our approach will be to show that, for the code to have rate
approaching capacity and error probability diminishing to
zero, then, with high probability,
the empirical distribution of $\bX^n$ needs to be near the set
of capacity-achieving input distributions. Since $\Vm = \Vmx$,
if the empirical distribution of $\bX^n$ is nearly capacity-achieving,
then the sum of the conditional variances of $\ii^*(X_k,Y_k)$ given
the past is
close to $n\Vm$ a.s., and a martingale central limit theorem\cite{Bolthausen}
can be applied. We begin with a few definitions needed for the
reduction to codes with empirically-capacity-achieving $\bX^n$.

\begin{definition}
The \emph{type} of a sequence $\bx^n$ is the distribution $P_{\bx^n}$ on $\X$ defined as 
\begin{align*}
P_{\bx^n}(a)\de\frac{1}{n}\sum_{k=1}^n\1\{x_k=a\}.
\end{align*} 
\end{definition}

\begin{definition}
For a sequence $\bx^n\in\X^n$, 
\begin{align*}
\phi_W(\bx^n)\de \inf_{P\in \Pi^\ast_W}d_{\text{TV}}(P,P_{\bx^n}),
\end{align*}
where $\dtv(P,Q)$ denotes the total variation distance between distributions $P$ and $Q$.
\end{definition}
\begin{definition}
Let $\mathcal{T}^n$ denote the set of all probability distributions on $\X$ that are types of some length-$n$ sequence, and define
\begin{align*}
\mathcal{T}^n_\gamma \de \left\{T\in \mathcal{T}^n, \inf_{P\in \Pi^\ast_W}d_{\text{TV}}(P,T)> \gamma\right\},\\
\mathcal{T}^{c,n}_\gamma \de \left\{T\in \mathcal{T}^n, \inf_{P\in \Pi^\ast_W}d_{\text{TV}}(P,T)\le \gamma\right\}.
\end{align*}
\end{definition}

Let $\Bf(m,\by^i)\de[f(m,\by^0), f(m,\by^1),\dots,f(m,\by^i)]\in\X^{i+1}$ with the convention that both $\by^0$  and $\Bf(m,\by^i)$ for $i\le -1$  are  empty strings.
\begin{definition}
If $Q$ is a probability distribution on $\X$ and $A \subset \X$ is
such that $Q(A) > 0$, then $Q|_A$ is the probability measure
\begin{equation}
Q_A(x) = \begin{cases}
\frac{Q(x)}{Q(A)} & \text{if $x \in A$} \\
0 & \text{otherwise}.
\end{cases}
\end{equation}
\end{definition}

\begin{definition}
Given a controller $F : (\X \times \Y)^* \mapsto \PP(\X)$, the
$(*,\gamma)$-modified controller $\tilde{F}$ is defined as follows.
For $k < n$ and $x^k \in \X^k$, let 
\begin{equation}
\X_{x^k} = \{x : (x^k,x) \ \text{is a prefix of some $x^n \in \mathcal{T}^{c,n}_\gamma$} \}.
\end{equation}
Fix some $x_0 \in \X$ arbitrarily. Let $\tilde{F}(x^k,y^k)$ be a 
point-mass on $x_0$ if either $k \ge n$ or $k < n$ but 
$F(x^k,y^k)(\X_{x^k}) = 0$ (note that the latter includes the
case in which $\X_{x^k}$ is empty). Otherwise, let
\begin{equation}
\tilde{F}(x^k,y^k) = F(x^k,y^k)|_{\X_{x^k}}.
\end{equation}
\end{definition}

\begin{definition}
Given a controller $F : (\X \times \Y)^* \mapsto \PP(\X)$, the
$(T,\gamma)$-modified controller is defined as in the previous
definition but with the type $T$ in place of 
$\mathcal{T}^{c,n}_\gamma$.
\end{definition}

 Lemma~\ref{thm:gen_conv_code} in the Appendix states for any $\rho_n>0$
\begin{align}
\log M_{\textnormal{fb}}^\ast(n, \varepsilon) \le \sup_{F} \inf_{q} \left(\log \rho_n-\log \left(\left[1-\varepsilon - F\circ W\left(\log \frac{\prod_{k=1}^n W(Y_k|X_k)}{q(\bY^n)} \ge \log \rho_n \right)\right]^+\right)\right),
\label{EQ:gen_conv}
\end{align}
where $F$ is a controller: $F:(\X \times \Y)^\ast \rightarrow \PP(\X)$. 
Let $P$ denote the distribution $F\circ W$. 
We will choose
\begin{align}
q(\by^n)=\frac{1}{2}\prod_{k=1}^n q^\ast(y_k) + \frac{1}{2|\mathcal{T}^n_\gamma|}\sum_{T\in\mathcal{T}^n_\gamma}\prod_{k=1}^n q_T(y_k),
\label{EQ:q_yn-def}
\end{align}
where
\begin{align*}
q_T(y)\de\sum_{x\in\X}T(x)W(y|x).
\end{align*}
This choice is inspired by an analogous choice by Fong and 
Tan~\cite[(37)]{Fong-Tan}, who in turn credit Hayashi~\cite{Hayashi}.

Let $K_W\de\max\left(2|\X|\nmx, \frac{8|\X|\nmx}{\Vm}\right)$ and $\chi_W$ denote the constant in~\cite[Corollary to Theorem 2]{Bolthausen} when
$\gamma$ in that result is taken to be $2\imx$ here. Fix $0<\gamma\le \frac{\Vm}{4|\X|\nmx}$, and define
\begin{align}
\delta_n &\de\chi_{W} \cdot \left(\frac{\log n}{\sqrt{n}(\Vm-\gamma K_{W})^{3/2}}+\sqrt{\gamma K_{W}}\right),\nn
r_n &\de \begin{cases}
\sqrt{\Vm-\gamma K_{W}}\Phi^{-1}\left(\varepsilon+3\delta_n\right)+\frac{\log 2}{\sqrt{n}} &\varepsilon\in\left(0,\frac{1}{2}-3\delta_n\right],\\
\sqrt{\Vm+\gamma K_{W}}\Phi^{-1}\left(\varepsilon+3\delta_n\right)+\frac{\log 2}{\sqrt{n}}&\varepsilon\in\left(\frac{1}{2}-3\delta_n,1\right),\\
\end{cases}\label{EQ:def-r_n}\\
\rho_n &\de\exp(nC + \sqrt{n}r_n)\nonumber.
\end{align}

We now analyze the probability term in~(\ref{EQ:gen_conv}). 
\begin{align}
P\left(\log \frac{\prod_{k=1}^n W(Y_k|X_k)}{q(\bY^n)}\ge \log \rho_n \right)&=P\left(\log \frac{\prod_{k=1}^n W(Y_k|X_k)}{q(\bY^n)} \ge \log \rho_n \bigcap \phi_W(\bX^n)\le\gamma\right)\nn
&\quad+P\left(\log \frac{\prod_{k=1}^n W(Y_k|X_k)}{q(\bY^n)} \ge \log \rho_n \bigcap \phi_W(\bX^n)>\gamma\right)\nn
&=P\left(\log \frac{\prod_{k=1}^n W(Y_k|X_k)}{q(\bY^n)} \ge \log \rho_n \bigcap \phi_W(\bX^n)\le\gamma\right)\nn
&\quad+\sum_{T\in \mathcal{T}^n_\gamma}P\left(\log \frac{\prod_{k=1}^n W(Y_k|X_k)}{q(\bY^n)} \ge \log \rho_n \bigcap P_{\bX^n} =T  \right).
\label{EQ:vmn_vmx_conv1}
\end{align}
We will now apply the code modification technique of Fong and Tan~\cite{Fong-Tan}.
Let $P_\ast$ (resp.\ $P_T$) denote the distribution induced by the $(\ast,\gamma)$-modified (resp.\ $(T,\gamma)$-modified)  code.
\begin{Lemma}
\label{Le:modified_code}
For an event $\mathcal{E}\in\sigma(\bX^n,\bY^n)$
\begin{align*}
P\left(\mathcal{E} \bigcap \phi_W(\bX^n)\le\gamma\right)\le P_\ast(\mathcal{E}),\\
P\left(\mathcal{E} \bigcap P_{\bX^n}=T\right)\le P_T(\mathcal{E}).
\end{align*}

\end{Lemma}
\begin{IEEEproof}
For any $(\bx^n,\by^n)$ such that $\phi_W(\bx^n) \le \gamma$, 
\begin{align}
P_*((\bx^n,\by^n)) & = \prod_{k = 1}^n \tilde{F} (x_k|\bx^{k-1},\by^{k-1})
                                            W(y_k|x_k) \\
             & = \prod_{k = 1}^n \frac{F (x_k|\bx^{k-1},\by^{k-1})
                                            W(y_k|x_k)}{F(\X_{x^{k-1}}|
                                      \bx^{k-1},\by^{k-1})} \\
     & \ge \prod_{k = 1}^n F(x_k|\bx^{k-1},\by^{k-1}) W(y_k|x_k) \\
     & = P(\bx^n,\by^n).
\end{align}
The proof of the second part is analogous.
\end{IEEEproof}
Application of the above lemma to~(\ref{EQ:vmn_vmx_conv1}) yields
\begin{align}
P\left(\log \frac{\prod_{k=1}^n W(Y_k|X_k)}{q(\bY^n)}\ge \log \rho_n \right)&\le
P_\ast\left(\log \frac{\prod_{k=1}^n W(Y_k|X_k)}{q(\bY^n)} \ge \log \rho_n\right)\nn
&\quad+\sum_{T\in \mathcal{T}^n_\gamma}P_T\left(\log \frac{\prod_{k=1}^n W(Y_k|X_k)}{q(\bY^n)} \ge \log \rho_n \right).
\label{EQ:vmn_vmx_conv3}
\end{align}

We will now upper bound the first term on the right-hand side of the above equation using a martingale central limit theorem. Let $\F_k=\sigma(M,Y_1,\dots,Y_k)$, and 
\begin{align}
Z_k&\de\ii^*(X_k,Y_k)-\E_\ast[\ii^*(X_k,Y_k)|\F_{k-1}]\label{EQ:Z-def},\\
S_k &\de \sum_{j=1}^k Z_j. \nonumber
\end{align}

\begin{align}
P_\ast\left(\log \frac{\prod_{k=1}^n W(Y_k|X_k)}{q(\bY^n)} \ge \log \rho_n \right) 
&\overset{(a)}{\le} P_\ast\left(\log \frac{\prod_{k=1}^n W(Y_k|X_k)}{1/2\prod_{k=1}^n q^\ast(Y_k)} \ge \log \rho_n \right)\nn
& =  P_\ast\left(\sum_{k=1}^n\left(\log \frac{W(Y_k|X_k)}{ q^\ast(Y_k)}-C\right) \ge \sqrt{n}r_n -\log 2 \right)\nn
&\overset{(b)}{=} P_\ast\left(\sum_{k=1}^n\left(\ii^*(X_k,Y_k)-\E_\ast[\ii^*(X_k,Y_k)|\F_{k-1}]\right) \ge \sqrt{n}r_n -\log 2 \right)\nn
&=P_\ast\left(\sum_{k=1}^n Z_k\ge \sqrt{n}r_n -\log 2 \right) \label{EQ:vmn-vmax-2},
\end{align}
where in (a), we have used the definition of $q(\bY^n)$ in (\ref{EQ:q_yn-def}), and \\*
 in (b), we  have used the fact that $\E_\ast[\ii^*(X_k,Y_k)|X_k]$  $=\sum_{y\in\Y}W(y|X_k)\log\frac{W(y|X_k)}{Q^*(Y_k)}\le C$~\cite[Theorem 4.5.1]{Gallager}.
\begin{Lemma}
\label{Le:*-code-var}
Let $\G_k=\sigma(S_1,\dots,S_k)$ for $1\le k \le n$, with $\G_0$ being the trivial $\sigma$-algebra. Then with $K_W=\max\left(2|\X|\nmx, \frac{8|\X|\nmx}{\Vm}\right)$,
\begin{align*}
\Vm-\gamma K_{W}\le\frac{1}{n}\sum_{k=1}^n\E_\ast[Z^2_k|\G_{k-1}]\le \Vm + \gamma K_{W}, \quad P_\ast\text{-}\as\\
\left\|\frac{\sum_{k=1}^n\E_\ast[Z^2_k|\G_{k-1}]}{\sum_{k=1}^n\E_\ast[Z^2_k]} -1\right\|_\infty \le \gamma K_{W},  \quad P_\ast\text{-}\as
\end{align*}
\end{Lemma}

\begin{IEEEproof}
The following chain of equalities holds  $P_\ast\text{-}\as$,
\begin{align*}
\frac{1}{n}\sum_{k=1}^n\E_\ast[Z^2_k|\F_{k-1}]&=\frac{1}{n}\sum_{k=1}^n\E_\ast[Z^2_k|X_k]\\
&=\frac{1}{n}\sum_{k=1}^n\V[\ii(X_k,Y_k)|X_k]\\
&=\frac{1}{n}\sum_{k=1}^n\sum_{x\in\X}\1\{X_k=x\}\nu_x\\
& = \sum_{x\in\X}P_{\bX^n}(x)\nu_x.
\end{align*}
Since $\phi_W(\bX^n)\le\gamma$, there exists a $\tilde{P}\in\Pi_W^\ast$ such that $d_{\text{TV}}(\tilde{P},P_{\bX^n})\le 2\gamma$. Thus we have for each $x\in\X$
\begin{align*}
|\tilde{P}(x)-P_{\bX^n}(x)|\le d_{\text{TV}}(\tilde{P},P_{\bX^n}) \le 2\gamma.
\end{align*}
Thus
\begin{align*}
\frac{1}{n}\sum_{k=1}^n\E_\ast[Z^2_k|\F_{k-1}] &= \sum_{x\in\X}P_{\bX^n}(x)\nu_x\nn
&\le  \sum_{x\in\X}\left(\tilde{P}(x) + 2\gamma\right)\nu_x\nn
&=\sum_{x\in\X}\tilde{P}(x)\nu_x+ 2\gamma\sum_{x\in\X}\nu_x\nn
&\le \Vm + 2\gamma|\X|\nmx,
\end{align*}
where the last step follows since for any $\tilde{P}\in\Pi_W^\ast$, $\sum_{x\in\X}\tilde{P}(x)\nu_x=\Vm$.

Similarly
\begin{align*}
\frac{1}{n}\sum_{k=1}^n\E_\ast[Z^2_k|\F_{k-1}] &\ge \Vm - 2\gamma|\X|\nmx.
\end{align*}
Since $\G_{k-1}\subseteq\F_{k-1}$, taking the conditional expectation with respect to $\G_{k-1}$, we get,
\begin{align*}
\Vm-2\gamma|\X|\nmx\le\frac{1}{n}\sum_{k=1}^n\E_\ast[Z^2_k|\G_{k-1}]\le \Vm + 2\gamma|\X|\nmx
\end{align*}
To prove the second part, we note that $P_\ast\text{-}\as$,
\begin{align*}
\left|\frac{\sum_{k=1}^n\E_\ast[Z^2_k|\G_{k-1}]}{\sum_{k=1}^n\E_\ast[Z^2_k]} -1\right| &\le \left|\frac{\Vm+2\gamma|\X|\nmx}{\Vm-2\gamma|\X|\nmx} -1\right|\\
&=\frac{4\gamma|\X|\nmx}{\Vm-2\gamma|\X|\nmx}\\
&\le \frac{8\gamma|\X|\nmx}{\Vm},
\end{align*}
provided $\gamma\le\frac{\Vm}{4|\X|\nmx}$. 

The statement of the lemma now follows since $K_W=\max\left(2|\X|\nmx, \frac{8|\X|\nmx}{\Vm}\right)$.
\end{IEEEproof}
Continuing the chain of expressions in (\ref{EQ:vmn-vmax-2}),
\begin{align}
P_\ast\left(\log \frac{\prod_{k=1}^n W(Y_k|X_k)}{q(\bY^n)} \ge \log \rho_n \right)
&\overset{}{\le} P_\ast\left(\sum_{k=1}^n Z_k\ge \sqrt{n}r_n -\log 2\right),\nn
&\overset{(a)}{\le} P_\ast\left(\frac{1}{\sqrt{\sum_{k=1}^n \E_\ast[Z^2_k]}}\sum_{k=1}^n Z_k\ge \Phi^{-1}(\varepsilon+3\delta_n) \right)\nn
&\overset{(b)}{\le} 1-\varepsilon-3\delta_{n}+\chi_W \cdot \left(\frac{n\log n}{\left(\sum_{k=1}^n \E_\ast[Z^2_k]\right)^{3/2}}+\left\| \frac{\sum_{k=1}^n\E_\ast[Z^2_k|\G_{k-1}]}{\sum_{k=1}^n\E_\ast[Z^2_k]}-1\right\|_\infty^{1/2}\right)\nn
&\overset{(c)}{\le} 1-\varepsilon-3\delta_{n}+\chi_W \cdot \left(\frac{\log n}{\sqrt{n}(\Vm-\gamma K_{W})^{3/2}}+\sqrt{\gamma K_{W}}\right)\nn
&= 1-\varepsilon-2\delta_n, \label{EQ:first-term}
\end{align}
where, for (a) we have used $n(\Vm-\gamma K_{W})\le\sum_{k=1}^n \E_\ast[Z^2_k]\le n(\Vm+\gamma K_{W})$ from Lemma~\ref{Le:*-code-var},\\*
for (b), we have used the martingale central limit theorem \cite[Corollary to Theorem 2]{Bolthausen}, taking the constant as $\chi_W$ (which only depends upon $\imx$ since $|Z_k|\le2\imx$ a.s.),\\*
for (c), we have used Lemma~\ref{Le:*-code-var}.

Moving to the second term in (\ref{EQ:vmn_vmx_conv3}), and noting that $q(\bY^n)\ge \frac{1}{2|\T_\gamma^n| }\prod_{k=1}^n q_T(Y_k)$, we get
\begin{align*}
\sum_{T\in \mathcal{T}^n_\gamma}P_T\left(\log \frac{\prod_{k=1}^n W(Y_k|X_k)}{q(\bY^n)} \ge \log \rho_n \right)
&\le \sum_{T\in \mathcal{T}^n_\gamma}P_T\left(\log \frac{\prod_{k=1}^n W(Y_k|X_k)}{\frac{1}{2|\T_\gamma^n| }\prod_{k=1}^n q_T(Y_k)} \ge \log \rho_n   \right)\\
&=\sum_{T\in \mathcal{T}^n_\gamma}P_T\left(\sum_{k=1}^n\log  \frac{ W(Y_k|X_k)}{ q_T(Y_k)} \ge \log \rho_n -\log 2|\T_\gamma^n|   \right).
\end{align*}
Consider
\begin{align*}
\sum_{k=1}^n\E_T\left[\log  \frac{ W(Y_k|X_k)}{ q_T(Y_k)}\middle|\F_{k-1}\right]&=\sum_{x\in\X}\sum_{k=1}^n\E_T\left[\log  \frac{ W(Y_k|X_k)}{ q_T(Y_k)}\middle|X_k=x\right]\1\{X_k=x\}\\
&=\sum_{x\in\X}\sum_{k=1}^n\sum_{y\in\Y}W(y|x)\log\frac{W(y|x)}{q_T(y)}\1\{X_k=x\}\\
&=n\sum_{x\in\X} T(x)\sum_{y\in\Y}W(y|x)\log\frac{W(y|x)}{q_T(y)}\\
&=nI(T;W).
\end{align*}
Recall that for any $P\in\Pi_W^*$ and $T\in \mathcal{T}^n_\gamma$, $\dtv(P,T)>\gamma>0$, hence  $I(T;W)<C$. Let $K_{T} \de C-I(T;W)>0$, and $\tilde{i}_{\text{max},T}\de\max_{x,y: W(y|x)q_T(y) > 0}\left|\log\frac{ W(y|x)}{ q_T(y)}\right|$. 

We now show that $\tilde{i}_{\text{max},T}\le 2 \log n$ $P_T$-a.s., for all sufficiently large $n$. Let $W_{\min} \de \min_{x,y:W(y|x)>0} W(y|x)$ and $q_{T,\min}\de\min_{q_T(y)>0}q_T(y)$. Then
\begin{align*}
q_{T,\min}\de\min_{q_T(y)>0}\sum_{x}T(x)W(y|x) \ge \min_{x,y:W(y|x)>0} W(y|x)\min_{x:T(x)> 0}T(x)
&=\frac{W_{\min}}{n},
\end{align*}
where the last equality follows since $T$ is the type of a sequence. 
Thus
\begin{align*}
\tilde{i}_{\text{max},T}&=\max_{x,y: W(y|x)q_T(y)>0}\left|\log\frac{ W(y|x)}{ q_T(y)}\right|\\
&\le \max_{x,y: W(y|x)q_T(y)>0}|\log W(y|x)|+\max_{y:q_T(y)>0} |\log { q_T(y)}|\\
&\le |\log W_{\min}|+\left|\log\frac{W_{\min}}{n}\right|\\
&=\log \frac{n}{W_{\min}^2}\\
&\le 2\log n
\end{align*}
for all sufficiently large $n$.

Defining $\tilde{Z_k}\de\log  \frac{ W(Y_k|X_k)}{ q_T(Y_k)}-E_T\left[\log  \frac{ W(Y_k|X_k)}{ q_T(Y_k)}\middle|\F_{k-1}\right]$,  we have
\begin{align}
&\sum_{T\in \mathcal{T}^n_\gamma}P_T\left(\sum_{k=1}^n\log  \frac{ W(Y_k|X_k)}{ q_T(Y_k)} \ge \log \rho_n -\log 2|\T_\gamma^n|   \right)\nn
&=\sum_{T\in \mathcal{T}^n_\gamma}P_T\left(\sum_{k=1}^n\left(\log  \frac{ W(Y_k|X_k)}{ q_T(Y_k)}-E_T\left[\log  \frac{ W(Y_k|X_k)}{ q_T(Y_k)}\middle|F_{k-1}\right]\right) \ge nK_{T}+\sqrt{n}r_n -\log 2|\T_\gamma^n|   \right)\nn
&=\sum_{T\in \mathcal{T}^n_\gamma}P_T \left(\sum_{k=1}^n \tilde{Z}_k \ge nK_{T}+\sqrt{n}r_n-\log 2|\T_\gamma^n| \right)\nn
&\overset{(a)}{\le}\sum_{T\in \mathcal{T}^n_\gamma}P_T\left(\sum_{k=1}^n \tilde{Z}_k \ge nK_{T}+\sqrt{n}r_n-|\X|\log 2(n+1)\right)\nn
&\overset{(b)}{\le}\sum_{T\in \mathcal{T}^n_\gamma} P_T \left(\sum_{k=1}^n \tilde{Z}_k \ge  \frac{nK_{T}}{2} \right)\nn
&\overset{(c)}{\le}\sum_{T\in \mathcal{T}^n_\gamma} \exp\left(-\frac{nK_{T}^2}{128\log^2 n} \right)\nn
&\overset{(d)}{\le}\sum_{T\in \mathcal{T}^n_\gamma} \exp\left(-\frac{nK}{\log^2 n} \right)\nn
&=  |\mathcal{T}_\gamma^n|\exp\left(-\frac{nK}{\log^2 n} \right)\nn
&\le (n+1)^{|\X|}\exp\left(-\frac{nK}{\log^2 n} \right)\nn
&\overset{(e)}{\le}\delta_n, \label{EQ:second-term}
\end{align}
where, (a) follows since $ |\T_\gamma^n|\le |\T^n| \le (n+1)^{|\X|}$,\\*
(b) follows since $\sqrt{n}r_n-|\X|\log 2(n+1)\ge-\frac{nK_{T}}{2}$ for all sufficiently large $n$,\\*
(c) follows from  Azuma's inequality~\cite[(3.3), p. 61]{Bercu15}, and noting that $|\tilde{Z}_k|\le 2\tilde{i}_{\text{max},T}\le 4 \log n$, \\*
(d) follows from defining $K\de\min_{T\in\mathcal{T}_\gamma^n} \frac{K_{T}^2}{128}$,\\*
(e) holds for all sufficiently large $n$.

From (\ref{EQ:vmn_vmx_conv3}), (\ref{EQ:first-term}), and (\ref{EQ:second-term}), we get
\[
P\left(\log \frac{\prod_{k=1}^n W(Y_k|X_k)}{q(\bY^n)}\ge \log \rho_n \right)\le 1-\varepsilon-\delta_n.
\]
Plugging the above inequality in (\ref{EQ:gen_conv}),
\begin{align*}
\log M_{\textnormal{fb}}^\ast(n, \varepsilon)\le \log \rho_n-\log \delta_n,
\end{align*}
i.e.,
\begin{align}
\label{eq:simpleoptfiniten}
\frac{\log M_{\textnormal{fb}}^\ast(n, \varepsilon)-nC}{\sqrt{n}}\le r_n-\frac{\log \delta_n}{\sqrt{n}}.
\end{align}
Using the definition of $r_n$ in (\ref{EQ:def-r_n}) and taking the limit
\begin{align*}
\limsup_{n\to\infty}\frac{\log M_{\textnormal{fb}}^\ast(n, \varepsilon)-nC}{\sqrt{n}}\le \begin{cases}
\sqrt{\Vm-\gamma K_{W}}\Phi^{-1}\left(\varepsilon+\chi_W\sqrt{\gamma K_W}\right) &\varepsilon\in\left(0,\frac{1}{2}-\chi_W \sqrt{\gamma K_W}\right],\\
\sqrt{\Vm+\gamma K_{W}}\Phi^{-1}\left(\varepsilon+\chi_W\sqrt{\gamma K_W}\right) &\varepsilon\in\left(\frac{1}{2}-\chi_W\sqrt{\gamma K_W},1\right).\\
\end{cases}
\end{align*}
Now taking $\gamma\to 0$ gives
\begin{align*}
\limsup_{n\to\infty}\frac{\log M_{\textnormal{fb}}^\ast(n, \varepsilon)-nC}{\sqrt{n}}\le \sqrt{\Vm}\Phi^{-1}\left(\varepsilon\right),
\end{align*}
proving the theorem.
\hfill\IEEEQED

\section{Proof of  Theorem~\ref{thm:conv_gen}}
\label{sec:proof_conv_gen}
We begin with a few definitions from stochastic calculus. Throughout we assume that the filtration under consideration is right-continuous and complete (via e.g.~\cite[Lemma 7.8, p. 124]{Kallenberg}).
\begin{definition}
A process $\N$ is called a \emph{local martingale} with respect to a filtration $(\F_t:t\ge 0)$ if $N_t$ is $\F_t$-measurable for each $t$ and there exists an increasing sequence of stopping times $T_n$, such that $T_n\to\infty$ and the stopped and shifted processes $\N^{T_n}\de (N_{\min\{t,T_n\}}-N_0:t\ge 0)$ are  $(\F_t:t\ge 0)$-martingales for each $n$.
\end{definition}

\begin{definition}
The \emph{quadratic variation} of a continuous local martingale $\N$ is an a.s.\ unique continuous process of locally finite variation, $[\N]$, such that $\N^2-[\N]$ is a local martingale. The existence and uniqueness of such process is guaranteed by~\cite[Theorem 17.5, p. 332]{Kallenberg}.
\end{definition}

\begin{definition}
A stochastic process is said to be $\F_t$-\emph{predictable}  if it is measurable with respect to the $\sigma$-algebra generated by all left-continuous $\F_t$-adapted processes. 
\end{definition}

By taking $q(\by^n)=\prod_{i=1}^n q^\ast(y_i)$ in~(\ref{eq:metaconverse}) in Lemma~\ref{thm:gen_conv_code} in the Appendix (which is almost
certainly a source of looseness in the bound), we get, for any $\rho_n > 0$,
\begin{align}
\log M_{\textnormal{fb}}^\ast(n, \varepsilon) \le \sup_{F} \left(\log \rho_n-\log \left(\left[1-\varepsilon - P\left(\sum_{k=1}^n \ii^*(X_k,Y_k) \ge \log \rho_n \right)^+\right]\right)\right),
\label{eq:conv_code_nmn}
\end{align}
where the supremum is over controllers: $F:(\X \times \Y)^\ast \rightarrow \PP(\X)$, and $P$ denotes the distribution $F\circ W$.
We use (\ref{eq:metaconverse}) over 
(\ref{eq:controlled:converse:hyp})-(\ref{eq:controlled:converse:conc}) 
in Lemma~\ref{thm:gen_conv_code} because it yields
a finite-$n$ result ((\ref{eq:finitenconverse}) to follow).
Fix an arbitrary $\kappa>0$, let $K_W \de 16\imx^2\nmx/\nmn$, and define
\begin{align}
\delta_n &\de \frac{ K_W}{\kappa^2\sqrt{n}},\\
r_n &\de 
\begin{cases}
\sqrt{\nmn}\Phi^{-1}\left(\frac{(1+\lambda)}{2\lambda}\left(\varepsilon+2\delta_n\right)\right)+\kappa, &  0<\varepsilon\le\frac{\lambda}{1+\lambda}-2\delta_n\\
\sqrt{\nmx}\Phi^{-1}\left( \frac{(\varepsilon+2\delta_n)(1+\lambda)+(1-\lambda)}{2}\right)+\kappa, & \frac{\lambda}{1+\lambda}-2\delta_n < \varepsilon < 1.
\end{cases}
\label{EQ:r_n}\\
\rho_n &\de\exp(nC + \sqrt{n}r_n).
\end{align}

The proof will consist of the following steps:
\begin{enumerate}
\item We will define a martingale sequence $(S_k, 1 \le k \le n)$ such that $P\left(\sum_{k=1}^n \ii^*(X_k,Y_k) \ge \log \rho_n \right) \le P(S_n \ge r_n)$.
\item We will embed the martingale sequence $(S_k, 1 \le k \le n)$ in a Brownian motion $\B$ such that  
$S_k=B_{T_k},  1 \le k \le n$, where $(T_k, 1 \le k \le n)$ are stopping times.
\item We will construct a process $\psi_t\in[\sqrt{\nmn}, \sqrt{\nmx}]$ and a Brownian motion $\W$ such that $\int_0^1 \psi_s\,dW_s\approx B_{T_n}$.
\item Applying a theorem from stochastic calculus, we will ``mimic'' the above It\^{o} process by a solution of a SDE $\hat{\boldsymbol{\xi}}$.
\item Using McNamara's result on the optimal control of diffusion processes~\cite{McNamara}, we will upper bound the probability $P\left(\hat{\xi}_1 \ge 0 \right)$ which will yield an upper bound on $P\left(\int_0^1 \psi_s\,dW_s \ge r_n \right)$.
\end{enumerate}
Proceeding, define 
\begin{align}
\F_k&\de\sigma(M,Y_1,\dots,Y_k),\nn
Z_k&\de\frac{1}{\sqrt{n}}\left(\ii^*(X_k,Y_k)-\E[\ii^*(X_k,Y_k)|\F_{k-1}]\right)\nn
S_k &\de \sum_{j=1}^k Z_j, \nn
\G_k&\de\sigma(S_1,\dots,S_k)\nonumber
\end{align}
We note that
\begin{align}
|Z_k|\le \frac{2}{\sqrt{n}}\imx \quad P-\mathrm{a.s.}
\label{EQ:Z-UB}
\end{align}

\begin{Lemma}
\label{Le:step-1}
 The sequence $(S_k, 1\le k \le n)$ is a martingale with respect to the filtration $(\G_k, 1\le k \le n)$ such that 
\begin{equation}
\label{EQ:Zvar}
\E[Z^2_k|\G_{k-1}]\in\left[\frac{\nmn}{n}, \frac{\nmx}{n} \right],
\end{equation}
and
\[
P\left(\sum_{k=1}^n \ii^*(X_k,Y_k) \ge \log \rho_n \right)\le P(S_n \ge r_n).
\]
\end{Lemma}
\begin{IEEEproof}

Since $\G_k\subseteq\F_k$ and  
\begin{align*}
\E[Z_k|\F_{k-1}]=0,
\end{align*}
taking the conditional expectation with respect to $\G_{k-1}$, we get
\begin{align*}
\E[Z_k|\G_{k-1}]=0.
\end{align*}
Thus the sequence $(S_k, 1\le k \le n)$ is a martingale with respect to the filtration $(\G_k, 1\le k \le n)$. 
Moreover
\begin{align}
\E[Z^2_k|\F_{k-1}]=\frac{1}{n}\sum_{x\in\X}\1\{X_k=x\}\nu_x\in\left[\frac{\nmn}{n}, \frac{\nmx}{n} \right].
\end{align}
Once again taking the conditional expectation with respect to $\G_{k-1}$, we get
\begin{align}
\E[Z^2_k|\G_{k-1}]\in\left[\frac{\nmn}{n}, \frac{\nmx}{n} \right].
\label{EQ:Z-range}
\end{align}
Now consider 
\begin{align}
P\left(\sum_{k=1}^n \ii^*(X_k,Y_k) \ge \log \rho_n \right)&=P\left(\frac{1}{\sqrt{n}}\sum_{k=1}^n (\ii^*(X_k,Y_k)-C) \ge  r_n\right)\nn
&\le P\left(\frac{1}{\sqrt{n}}\sum_{k=1}^n \left(\ii^*(X_k,Y_k)-\E[\ii^*(X_k,Y_k)|\F_{k-1}]\right) \ge r_n \right)\nn
&= P(S_n \ge r_n),
\end{align}
where in the middle step we  have used the fact that~\cite[Theorem 4.5.1]{Gallager} 
$$\E[\ii^*(X_k,Y_k)|\F_{k-1}]=\E[\ii^*(X_k,Y_k)|X_k]=\sum_{y\in\Y}W(y|X_k)\log\frac{W(y|X_k)}{Q^*(Y_k)}\le C.$$
\end{IEEEproof}

\begin{Lemma}
There exists a Brownian motion $\B$, and a sequence of non-decreasing stopping times $T_1,\dots, T_n$ such that
\begin{align*}
S_k = B_{T_k} \,\,\,\text{a.s.} \qquad k\in\{1,\dots,n\},
\end{align*}
and if $\tilde{\G}_k=\sigma(S_1,T_1\dots,S_k,T_k)$, and $\tau_k=T_k-T_{k-1}$ (with $T_0 = 0$), then
\begin{align}
E[\tau_k|\tilde{\G}_{k-1}] = \E[Z_k^2|\G_{k-1}],\label{EQ:tau}\\
E[\tau_k^2|\tilde{\G}_{k-1}] \le 4\E[Z_k^4|\G_{k-1}] \label{EQ:tau1}.
\end{align}
\end{Lemma}
\begin{IEEEproof}
The lemma is a straightforward application of~\cite[Theorem 14.16, p. 279]{Kallenberg} to the martingale sequence $(S_k, 1\le k \le n)$.
\end{IEEEproof}

\begin{Lemma}
\label{Le:1}
There exists a filtration $\Hc_t$,  an $\Hc_t$-predictable process $\boldsymbol{\psi}$, an $\Hc_t$ Brownian motion $\W$, and an $\Hc_t$-stopping time $T_n^\ast$ such that
\begin{enumerate}
\item$\sqrt{\nu}_\text{min} \le \psi_t \le \sqrt{\nu}_\text{max}$ a.s.
\item$
\int_0^{T_n^\ast} \psi_t \,dW_t=B_{T_n}=S_n.
$
\item $\E[(T_n^\ast-1)^2] \le \frac{K_W^{(1)}}{n} $, where $K_W^{(1)}\de {64\imx^4}/{\nmn^2}$.
\end{enumerate}
\end{Lemma}

\begin{IEEEproof}
Define $\boldsymbol{\psi}$ as
\begin{align}
\psi_t=
\begin{cases}
\sqrt{n\E[\tau_1|\tilde{\G}_{0}]} &\quad 0\le t \le \frac{\tau_1}{n\E[\tau_1|\tilde{\G}_{0}]} \\
\sqrt{n\E[\tau_2|\tilde{\G}_{1}]} &\quad\frac{\tau_1}{n\E[\tau_1|\tilde{\G}_{0}]}< t \le \frac{\tau_1}{n\E[\tau_1|\tilde{\G}_{0}]}+\frac{\tau_2}{n\E[\tau_2|\tilde{\G}_{1}]} \\
\qquad\vdots & \quad \qquad\qquad\vdots\\
\sqrt{n\E[\tau_n|\tilde{\G}_{n-1}]} &\quad\sum_{j=1}^{n-1}\frac{\tau_j}{n\E[\tau_j|{\tilde{\G}}_{j-1}]}< t \le \sum_{j=1}^{n}\frac{\tau_j}{n\E[\tau_j|\tilde{\G}_{j-1}]} \\
\sqrt{\nu}_\text{min} &\quad t> \sum_{j=1}^{n}\frac{\tau_j}{n\E[\tau_j|\tilde{\G}_{j-1}]}
\end{cases}
\label{EQ:psi}
\end{align}

Then, from the above definition, (\ref{EQ:tau}), and (\ref{EQ:Zvar}), it is clear that $\sqrt{\nu}_\text{min} \le \psi_t \le \sqrt{\nu}_\text{max}$ a.s.

We now  employ the change-of-time method (see~\cite{Barndorff15}). To illustrate the reason behind it, consider the stochastic integral
\[
\tilde{\xi}_t =  \int_{0}^t\tilde{\psi}_s\,d\tilde{W}_s,
\]
with $\tilde{\W}$ being a Brownian motion and $\tilde{\psi}_s\in[\nmn,\nmx]$ being a predictable step process. Let $\tilde{A}_t\de[\tilde{\xi}]_t=\int_{0}^t{\psi}^2_s\,ds$~\cite[Lemma 17.10 and Theorem 18.3]{Kallenberg}. Moreover,  $\tilde{\xi}_t=\tilde{B}_{\tilde{A}_t}$ for some Brownian motion $\tilde{\B}$~\cite[Theorem 18.4, p. 352]{Kallenberg}. Let $\tilde{T}\de\tilde{A}_1$, then 
\[
B_{\tilde{T}}=B_{\tilde{A}_1}=\tilde{\xi}_1=\int_{0}^1\tilde{\psi}_s\,d\tilde{W}_s.
\]

Hence, if by choosing $\bold{\tilde{\xi}}$ properly,  we could ensure that $\tilde{T}=T_n$ and $\tilde{\B}=\B$, then we would have proven a stronger version of the  lemma (with $T_n^\ast=1$). However, proving this stronger result appears to be difficult, and hence we allow  $T_n^\ast$ to be random. We continue with the proof of the lemma. 

Let $A_t\de\int_0^{t}\psi^2_s\,ds$. We note that  $\A$ is continuous   and strictly increasing, and we define the following time changed process 
$\mathbf{N}\de \mathbf{B}\circ\mathbf{A}$, i.e., $$N_t=B_{A_t}=B_{\int_0^{t}\psi^2_s\,ds},$$ and
\begin{align*}
\mathcal{H}_t\de \sigma({B}_{A_s}, 0 \le s \le t).
\end{align*}
Let $$T_k^\ast=\sum_{j=1}^{k}\frac{\tau_j}{n\E[\tau_j|\tilde{\G}_{j-1}]},\quad 1\le k \le n,$$ then it follows that (see Figure~\ref{Fig:A})
\begin{align*}
A_{T_k^\ast}=\int_0^{T_k^\ast}\psi^2_t\,dt =\sum_{j=1}^k \tau_j = T_k,\quad 1\le k \le n.
\end{align*}
Hence, $T_n^\ast=A^{-1}_{T_n}$, where $A^{-1}_t(\omega)$ is the inverse of $A_t(\omega)$ for each $\omega$ in the given sample space.  We can write
\[
T_n = \inf\{t>0;A^{-1}_t>T_n^\ast\} 
\]
 Noting that $A^{-1}_t$ is continuous and $T_n$ is a  $\sigma({B}_{s}, 0 \le s \le t)$-stopping time, applying~\cite[Proposition 7.9, p. 124]{Kallenberg}, we conclude that $A^{-1}_{T_k}=T_k^\ast$ is an $\mathcal{H}_t$-stopping time for each $k$ (the role of process $X_t$ in~\cite[Proposition 7.9, p. 124]{Kallenberg} is played by $A^{-1}_{t}$ here).

\begin{figure}
\begin{center}
\includegraphics[scale=0.15]{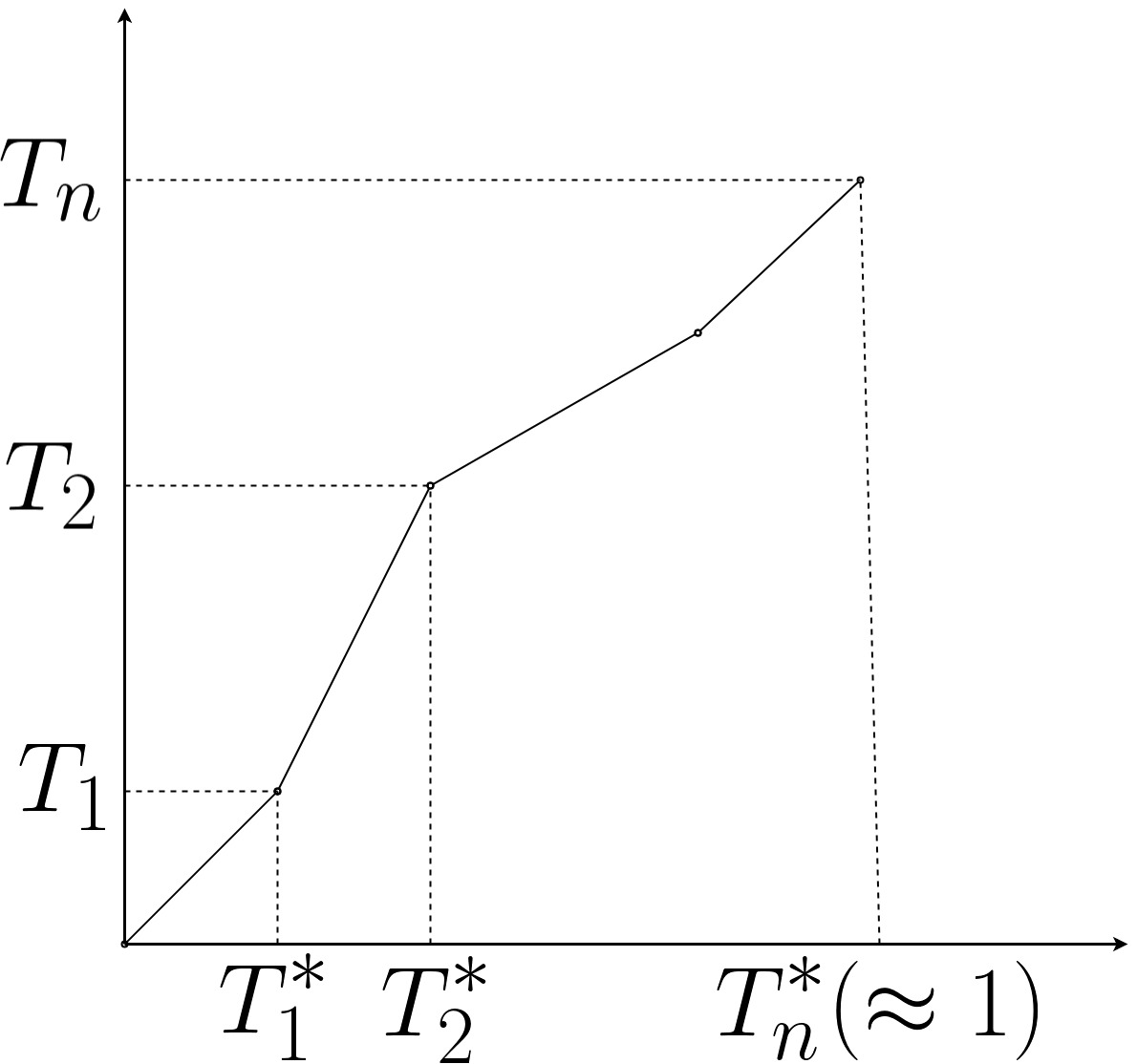}
\caption{Plot of $A_t$ vs $t$ for a fixed $\omega$ in the sample space.}
\label{Fig:A}
\end{center}
\end{figure}

Now  applying \cite[Theorem 17.24, p. 344]{Kallenberg} we get that $\N$ is a continuous  local martingale with respect to the filtration $\mathcal{H}_t$ with quadratic variation
\begin{align}
[\N]=[\B]\circ\mathbf{A}=\mathbf{A},
\end{align}
since $[B]_t = t$~\cite[Theorem 18.3, p. 352]{Kallenberg}. Now we follow the proof of \cite[Theorem 4.2, p. 170]{Karatzas-Shreve}. Define $\mathbf{W}$ as
\begin{align*}
W_t=\int_0^t\frac{1}{{\psi_s}}\,dN_s.
\end{align*}
Then $\mathbf{W}$ is a continuous local martingale with quadratic variation (\cite[Lemma 17.10, p.335]{Kallenberg}, noting that $1/\psi_s$ is a step process)
\begin{align*}
[W]_t = \int_0^t\frac{1}{{\psi^2_s}}\,d[N]_s=\int_0^t\frac{1}{{\psi^2_s}}\psi^2_s\,ds = t,
\end{align*} 
where we have used~\cite[Proposition 17.14, p. 338]{Kallenberg} for the middle equality. Hence $\mathbf{W}$ is a standard Brownian motion with respect to the filtration $\mathcal{H}_t$~\cite[Theorem 18.3, p. 352]{Kallenberg}.

Noting that there exists a (random) partition $0=t_0<t_1,\dots,<t_l=t$ such that $\boldsymbol{\psi}$ is constant on $(t_k, t_{k+1}]$ for $0\le k \le l-1$,  we can write
\begin{align*}
\int_0^t\psi_s\,dW_s =\sum_{k=0}^{l-1}\psi_{t_k}(W_{t_{k+1}}-W_{t_{k}})=\sum_{k=0}^{l-1}\psi_{t_k}\frac{1}{\psi_{t_k}}(N_{t_{k+1}}-N_{t_{k}})=N_t.
\end{align*} 

Thus 
\begin{align}
\int_0^{T_n^\ast}\psi_s\,dW_s=N_{T_n^\ast}=B_{A_{T_n^\ast}} = B_{T_n} =S_n.
\end{align}
Since $T_k^*$ is an $\Hc_t$ stopping time for each $k$, $\psi$ is
adapted to $\Hc_t$. Since is it left continuous, it is also predictable.

Now we bound $\E[(T_n^\ast-1)^2]$:
\begin{align*}
\E[(T_n^\ast-1)^2] &= \E\left[\left( \sum_{j=1}^{n}\frac{\tau_j}{n\E[\tau_j|\tilde{\G}_{j-1}]}-1\right)^2\right]\\
&=\E\left[\left( \sum_{j=1}^{n}\frac{\tau_j-\E[\tau_j|\tilde{\G}_{j-1}]}{n\E[\tau_j|\tilde{\G}_{j-1}]}\right)^2\right]\\
&\overset{(a)}{\le}\frac{1}{\nmn^2}\E\left[\left( \sum_{j=1}^{n}{\tau_j-\E[\tau_j|\tilde{\G}_{j-1}]}\right)^2\right]\\
&\overset{(b)}{=}\frac{1}{\nmn^2}\E\left[ \sum_{j=1}^{n}\left({\tau_j-\E[\tau_j|\tilde{\G}_{j-1}]}\right)^2\right]\\
&\overset{(c)}{\le}\frac{1}{\nmn^2}\E\left[ \sum_{j=1}^{n}\E[\tau^2_j|\tilde{\G}_{j-1}]\right]\\
&\overset{(d)}{\le}\frac{4}{\nmn^2}\E\left[ \sum_{j=1}^{n}\E[Z^4_j|\G_{j-1}]\right]\\
&\overset{(e)}{\le}\frac{4}{\nmn^2}\E\left[ \sum_{j=1}^{n}\frac{16\ii_{max}^4}{n^2}\right]\\
&= \frac{64\imx^4}{n\nmn^2}\\
&\overset{(f)}{=} \frac{K_W^{(1)}}{n}.
\end{align*}
Here,  (a) follows from (\ref{EQ:Z-range}) and (\ref{EQ:tau}),\\*
 (b) follows from noting that the sequence $(\tau_j-\E[\tau_j|\tilde{\G}_{j-1}], 1\le j \le n)$ is a martingale difference sequence with respect to the filtration $(\tilde{\G}_{j},1\le j \le n)  $, making $(\sum_{j=1}^k\tau_j-\E[\tau_j|\tilde{\G}_{j-1}], 1\le k \le n)$ a martingale and the orthogonal increment property of martingales~\cite[Theorem 5.4.6]{Durrett10}, \\*
 (c) follows from $\E[({\tau_j-\E[\tau_j|\tilde{\G}_{j-1}]})^2|\G_{j-1}]=\E[\tau^2_j|\tilde{\G}_{j-1}]-\left(\E[\tau_j|\tilde{\G}_{j-1}]\right)^2$,\\*
(d) follows from (\ref{EQ:tau1}),\\*
(e) follows since $|Z_j|\le \frac{2}{\sqrt{n}}\imx$ a.s. from (\ref{EQ:Z-UB}),\\*
(f) follows from defining $K_W^{(1)}\de {64\imx^4}/{\nmn^2}$.
\end{IEEEproof}
Now define
\begin{align}
{\xi}_t \de -(r_n-\kappa) + \int_0^t {\psi}_s\,d{W}_s.
\label{EQ:xi}
\end{align}
We have the following lemma.
\begin{Lemma}
\begin{align*}
P\left(\int_0^{T_n^\ast}\psi_s\,dW_s\ge r_n \right)\le P\left(\xi_1 \ge 0 \right)+\delta_n.
\end{align*}
\end{Lemma}
\begin{IEEEproof}

\begin{align*}
P\left(\int_0^{T_n^\ast}\psi_s\,dW_s\ge r_n \right)=P\left(\int_0^{1}\psi_s\,dW_s+\theta_n\ge r_n \right),
\end{align*}
where we have defined $\theta_n$ as
\begin{align*}
\theta_n \de \int_0^{\infty}\1\{1< s \le T_n^\ast\}\psi_s\,dW_s - \int_0^{\infty}\1\{T_n^\ast \le s < 1\}\psi_s\,dW_s.
\end{align*}
The second moment of $\theta_n$ can be bounded as
\begin{align*}
\E[\theta_n^2] &\overset{(a)}{\le} 2\E\left[\left(\int_0^{\infty}\1\{1< s \le T_n^\ast\}\psi_s\,dW_s \right)^2\right] + 2\E\left[\left(\int_0^{\infty}\1\{T_n^\ast \le s < 1\}\psi_s\,dW_s \right)^2\right]\\
&\overset{(b)}{=}2\E\left[\int_0^{\infty}\1\{1< s \le T_n^\ast\}\psi^2_s\,ds \right] + 2\E\left[\int_0^{\infty}\1\{T_n^\ast \le s < 1\}\psi^2_s\,ds \right]\\
&= 2\E\left[\1\{1< T_n^\ast\}\int_1^{T_n^\ast}\psi^2_s\,ds \right] + 2\E\left[\1\{T_n^\ast< 1\}\int^1_{T_n^\ast}\psi^2_s\,ds \right]\\
&\le 2\nmx\E[|T_n^\ast-1|]\\
&\le  2\nmx\sqrt{\E[(T_n^\ast-1)^2]}\\
&\overset{(c)}{\le} \frac{K_W}{\sqrt{n}}.
\end{align*}
Here, for (a) we have used the inequality $(a-b)^2\le 2a^2+2b^2$,\\*
for (b) we have used~\cite[Problem 2.18, p. 144]{Karatzas-Shreve},\\* 
for (c) we have used Lemma~\ref{Le:1}, and recalling $K_W=16\imx^2\nmx/\nmn=2\nmx\sqrt{K_W^{(1)}}$.

Thus
\begin{align*}
P\left(\int_0^{T_n^\ast}\psi_s\,dW_s\ge r_n \right)&=P\left(\int_0^{1}\psi_s\,dW_s+\theta_n\ge r_n \right)\\
&=P\left(\int_0^{1}\psi_s\,dW_s+\theta_n\ge r_n\bigcap |\theta_n|\le \kappa \right) + P\left(\int_0^{1}\psi_s\,dW_s+\theta_n\ge r_n \bigcap |\theta_n|> \kappa \right) \\
&\le P\left(\int_0^{1}\psi_s\,dW_s\ge r_n-\kappa\bigcap |\theta_n|\le \kappa \right) + P\left(\int_0^{1}\psi_s\,dW_s+\theta_n\ge r_n \bigcap |\theta_n|> \kappa \right)\\
&\le P\left(\int_0^{1}\psi_s\,dW_s\ge r_n-\kappa \right) + P\left(|\theta_n|> \kappa \right)\\
&\le P\left(\int_0^{1}\psi_s\,dW_s\ge r_n-\kappa \right) + \frac{E[\theta_n^2]}{\kappa^2}\\
&\le P\left(\int_0^{1}\psi_s\,dW_s\ge r_n-\kappa \right) + \frac{K_W}{\kappa^2\sqrt{n}}\\
&= P\left(\int_0^{1}\psi_s\,dW_s\ge r_n-\kappa \right) + \delta_n\\
& =  P\left(\xi_1 \ge 0 \right)+\delta_n. \qedhere
\end{align*}
\end{IEEEproof}
Now we apply~\cite[Corollary 3.7]{Brunick2013} (see also~\cite{Gyongy1986}). There exists a  probability space with a measure $\hat{P}$ that supports a process  $\hat{\boldsymbol{\xi}}$ and a  Brownian motion $\hat{\W}$ such that
\begin{align}
\hat{\xi}_t &= -(r_n-\kappa) + \int_0^t \hat{\psi}_s(\hat{\xi}_s)\,d\hat{W}_s,
\label{EQ:hat_xi}\\
 P\left(\xi_t \ge a \right) &=  \hat{P}\left(\hat{\xi}_t \ge a \right), \qquad a\in\R, \, t\ge0,
\end{align}
and $\hat{\psi}_t(\cdot)$ satisfies
\begin{align*}
\hat{\psi}^2_t(u)=\E[\psi^2_t|{\xi}_t=u] \qquad P\text{-}\as,\,t\in \mathcal{N}^c,
\end{align*}
where $\mathcal{N}$ is a Lebsegue-null set. In particular, we can take $\hat{\psi}_t(u)=\sqrt{\E[\psi^2_t|{\xi}_t=u]}$~\cite[Section 5.3]{Stroock2007}.

Note that  $\boldsymbol{\xi}$ in (\ref{EQ:xi}) is an It\^{o} process, where, in general the drift coefficient $\boldsymbol{\psi}$ itself can be a stochastic process. The process $\boldsymbol{\hat{\xi}}$, on the other hand, has deterministic function $\boldsymbol{\hat{\psi}}(\cdot)$ as the drift coefficient and the same one-dimensional law as that of $\boldsymbol{\xi}$ for each $t$.

Since $\hat{\psi}_t \in[\sqrt{\nmn}, \sqrt{\nmx}]$, (\ref{EQ:xi}) has a unique 
solution in distribution~\cite[Exercise 7.3.3]{Stroock2007} (see also the discussion after~\cite[Corollary 3.13]{Brunick2013}). Thus the setup in (\ref{EQ:hat_xi}) is \emph{admissible} as defined by McNamara in~\cite{McNamara}. McNamara~\cite[Remark 8]{McNamara}  shows that  if the goal is to maximize  $\hat{P}\left(\bar{\xi}_1 \ge 0 \right)$ where
\begin{align*}
\bar{\xi}_t = -(r_n-\kappa) + \int_0^t \bar{\psi}_s(\bar{\xi}_s)\,d\hat{W}_s,
\end{align*}
by choosing the optimal   diffusion coefficient $\bar{\psi}_s(\cdot)$, then such optimal diffusion control is given   by
\begin{align}
\bar{\psi}^{\text{opt}}(u)\de\sqrt{\nmn}\mathbf{1}\{u > 0\}+\sqrt{\nmx}\mathbf{1}\{u \le 0\}.
\label{EQ:ti_xi1}
\end{align}
Let the corresponding SDE be
\begin{align}
\bar{\xi}_t^{\text{opt}} \de -(r_n-\kappa) + \int_0^t \bar{\psi}^{\text{opt}}(\bar{\xi}_s^{\text{opt}})\,d\hat{W}_s.
\label{EQ:ti_xi2}
\end{align}
Thus
\begin{align}
\hat{P}\left(\hat{\xi}_1 \ge 0 \right)\le \hat{P}\left(\bar{\xi}_1^{\text{opt}}  \ge 0 \right).
\end{align}
Using the distribution function of the solution to  (\ref{EQ:ti_xi1}) and (\ref{EQ:ti_xi2}) (see~(\ref{eq:markov-xi-delta-0-trans})), we get
\begin{align}
\hat{P}\left(\bar{\xi}_1^{\text{opt}} \ge 0 \right) = 1-\frac{2\lambda}{1+\lambda}\Phi\left(\frac{r_n-\kappa}{\sqrt{\nmn}} \right),
\end{align}
when $r_n-\kappa \le 0$, and 
\begin{align}
\hat{P}\left(\bar{\xi}_1^{\text{opt}} \ge 0 \right) = \frac{2}{1+\lambda}-\frac{2}{1+\lambda}\Phi\left(\frac{r_n-\kappa}{\sqrt{\nmx}} \right),
\end{align}
when $r_n-\kappa > 0$. For our choice of $r_n$ in (\ref{EQ:r_n}), we get
\begin{align}
\hat{P}\left(\bar{\xi}_1^{\text{opt}} \ge 0 \right) = 1-\varepsilon-2\delta_n.
\end{align}
Summarizing the chain of inequalities so far, we have
\begin{align*}
P\left(\sum_{k=1}^n \ii^*(X_k,Y_k)\ge \log\rho_n \right) &\le P(S_n \ge r_n)\\
&=P\left(\int_0^{T_n^\ast}\psi_s\,dW_s\ge r_n \right)\\
&\le P\left(\xi_1 \ge 0 \right)+\delta_n \\
& = \hat{P}\left(\hat{\xi}_1 \ge 0 \right)+\delta_n \\
& \le \hat{P}\left(\bar{\xi}^{\text{opt}} _1 \ge 0 \right) + \delta_n \\
&= 1-\varepsilon-\delta_n.
\end{align*}
Thus from (\ref{eq:conv_code_nmn})
\begin{align}
\label{eq:finitenconverse}
\log M_{\textnormal{fb}}^\ast(n, \varepsilon) \le nC+\sqrt{n}r_n - \log \frac{ K_W}{\kappa^2\sqrt{n}},
\end{align}
and hence
\begin{align*}
\frac{\log M_{\textnormal{fb}}^\ast(n, \varepsilon) -nC}{\sqrt{n}} \le r_n -\frac{1}{\sqrt{n}} \log \frac{ K_W}{\kappa^2\sqrt{n}}.
\end{align*}
From the  definition of $r_n$ in~(\ref{EQ:r_n}), and taking $n\to\infty$,
\begin{equation*}
\limsup_{n \to \infty} \frac{\log M^\ast_{\textnormal{fb}}(n, \varepsilon) - n\mC}{\sqrt{n}}  \leq 
\begin{cases}
\sqrt{\nmn}\Phi^{-1}\left( \frac{1}{2 \lambda}\varepsilon(1+\lambda)\right)+\kappa, &  \varepsilon \in (0,\frac{\lambda}{1+\lambda}], \\
\sqrt{\nmx}\Phi^{-1}\left( \frac{1}{2}[\varepsilon(1+\lambda) + (1-\lambda)]\right)+\kappa, & \varepsilon \in (\frac{\lambda}{1+\lambda},1).
\end{cases}
\end{equation*}
Since $\kappa$ is arbitrary, we may take $\kappa\to 0$ to prove the theorem.
\hfill\IEEEQED

\section{Very Noisy Channels}
\label{sec:vnc}

We first derive the scaling behavior of various channel parameters ($C_\zeta$, $V_{\min,\zeta}$, etc.) with respect to $\zeta$.  
Recall that the VNC is given by
$$
W_\zeta(y|x)=\Gamma(y)\left(1+\zeta\lambda(x,y)\right),
$$
where $\Gamma$ is a probability distribution on the output alphabet $\Y$,
which we may assume, without loss of generality, has full support,
$\lambda(x,y)$ satisfies 
\begin{equation}
\label{EQ:Lambda_Sum2}
\sum_{y\in\Y}\Gamma(y)\lambda(x,y)=0
\end{equation}
for all $x\in\X$, and $\zeta$ is infinitesimally small. Let 
\begin{align*}
\lambda_{\text{max}}\de\max_{x\in\X,y\in\Y}|\lambda(x,y)|.
\end{align*}
We will denote by $K(\lm)$ any non-negative constant which depends only on $(\lambda_{\text{max}},|\X|,|\Y|)$. The quantity represented by $K(\lm)$ will
in general change from line to line in the derivation.

We will use the following approximation throughout the proof:
\begin{Lemma}
For all $u$ sufficiently close to zero,
\begin{align*}
|\log(1+u)-u|\le u^2, \nn
\left|\log(1+u)-\left(u-\frac{u^2}{2}\right)\right|\le u^3.
\end{align*}
\end{Lemma}

The following lemma gives the scaling of the capacity $C_\zeta$ of the above channel.
\begin{Lemma}
\label{Le:C_Scaling}
 Let $C_\zeta$ denote the capacity of $W_\zeta$. Then, for all sufficiently small $\zeta$,
\begin{align*}
\left|C_\zeta-\zeta^2\C\right|\le\zeta^3 K(\lm).
\end{align*}
where
\begin{align}
\label{EQ:VNCCdef}
\C\de\max_{P\in\mathcal{P}(\X)}\frac{1}{2}\sum_{y\in\Y}\Gamma(y)\Bigg(\left.\sum_{x\in \X}P(x)\lambda^2(x,y)-\left( \sum_{x\in\X}P(x)\lambda(x,y)\right)^2 \right).
\end{align}
\end{Lemma}
\begin{IEEEproof}
Let $\lambda_P(y)=\sum_{x\in\X}P(x)\lambda(x,y)$. The channel capacity at $\zeta$ is given by
\begin{align*}
C_\zeta&=\max_{P\in\mathcal{P}(\X)}I(P;W_\zeta)\nn
&=\max_{P\in\mathcal{P}(\X)}\sum_{x\in\X, y\in\Y}P(x)\Gamma(y)\left(1+\zeta\lambda(x,y)\right)\log\frac{1+\zeta\lambda(x,y)}{1+\zeta\lambda_P(y)}\nn
&\le\max_{P\in\mathcal{P}(\X)}\sum_{x\in\X, y\in\Y}P(x)\Gamma(y)\left(1+\zeta\lambda(x,y)\right)
\left(\zeta\lambda(x,y)-\frac{\zeta^2\lambda^2(x,y)}{2}-\zeta\lambda_P(y)  +\frac{\zeta^2\lambda_P^2(y)}{2}\right)+ \zeta^3K(\lm)\nn
&\overset{(a)}{\le}\max_{P\in\mathcal{P}(\X)}\sum_{y\in\Y}\Gamma(y)\left(\frac{\zeta^2}{2}\sum_{x\in\X}P(x)\lambda^2(x,y)  -\frac{\zeta^2}{2}\lambda^2_P(y)\right)  + \zeta^3K(\lm)\nn
&= \zeta^2 \C + \zeta^3K(\lm).
\end{align*}
Here for (a), we note that $\sum_{y\in\Y}\Gamma(y)\lambda(x,y)=0$, hence all the terms involving $\zeta$ disappear. The terms involving $\zeta^3$ have been absorbed in  $\zeta^3K(\lm)$. Similarly, we can show  $C_\zeta\ge\zeta^2 \C - \zeta^3K(\lm)$.
\end{IEEEproof}

Let $q_\zeta^\ast$ denote the  output distribution corresponding to a capacity-achieving input distribution $P_\zeta^*$, i.e.,
\begin{align*}
q_\zeta^\ast(y)=\Gamma(y)(1+\zeta\lambda^*_\zeta(y)),
\end{align*}
where
\begin{align*}
\lambda^*_\zeta(y)\de\sum_{x\in\X}P_\zeta^*(x)\lambda(x,y).
\end{align*}
Here, we note that $|\lambda^*_\zeta(y)|\le \lambda_{\text{max}}$, and 
\begin{align}
\sum_{y\in\Y}\Gamma(y)\lambda^*_\zeta(y)=0.
\label{EQ:lamba_prime_sum}
\end{align} 
 Also, since $q_\zeta^\ast$ is unique, $\lambda^*_\zeta$ is also unique.
Define
\begin{align*}
\X_\zeta^*\de\left\{x:\E[i^*(X,Y)|X=x]=C_\zeta\right\}.
\end{align*}
For $x\notin\X_\zeta^*$, let 
$$\rho_{\zeta,x}\de C_\zeta-\E[i^*(X,Y)|X=x],$$ 
where we note that
$\rho_{\zeta,x}>0$ \cite[Theorem 4.5.1]{Gallager}.

Define, for each $x\in\X$,
\begin{align*}
\nu_{x,\zeta}\de\text{Var}\left[i^*(X,Y)|X=x \right].
\end{align*}
\begin{Lemma}
\label{Le:IR_Scaling}
For all sufficiently small $\zeta$, the conditional expectation and variance of $i^*(X,Y)$ satisfy, for each $x$,
\begin{align*}
\left|\E[i^*(X,Y)|X=x]-\zeta^2\Psi_{\zeta,x}\right|\le\zeta^3 K(\lm),\nn
\left|\nu_{x,\zeta}-2\zeta^2\Psi_{\zeta,x}\right|\le\zeta^3 K(\lm),
\end{align*}
where
$$
\Psi_{\zeta,x}\de \frac{1}{2}\sum_{y\in\Y}\Gamma(y)\left(\lambda(x,y)-\lambda^*_\zeta(y) \right)^2.
$$
Hence for $x\in\X_\zeta^*$,
$$
\left|\nu_{x,\zeta}-2C_\zeta\right|\le\zeta^3 K(\lm),
$$
and for $x\notin\X_\zeta^*$,
$$
\left|\nu_{x,\zeta}-2(C_\zeta-\rho_{\zeta,x})\right|\le\zeta^3 K(\lm).
$$
\end{Lemma} 
\begin{IEEEproof}
We first note that since $|\lambda^*_\zeta(y)|\le\lambda_{\text{max}} $, we have  $\Psi_{\zeta,x}\le K(\lm)$.
Now consider,
\begin{align*}
\E[i^*(X,Y)|X=x]
&=\sum_{y\in\Y}W_\zeta(y|x)\log\frac{W_\zeta(y|x)}{q^\ast_\zeta(y)} \\
&=\sum_{y\in\Y}\Gamma(y)\left(1+\zeta\lambda(x,y)\right)\log\frac{1+\zeta\lambda(x,y)}{1+\zeta\lambda^*_\zeta(y)}\nn
&\le\sum_{y\in\Y}\Gamma(y)\left(1+\zeta\lambda(x,y)\right)\left(\zeta\lambda(x,y)-\frac{\zeta^2\lambda^2(x,y)}{2}-\zeta\lambda^*_\zeta(y) +\frac{\zeta^2\lambda^{* 2}_\zeta(y)}{2}  \right)+\zeta^3K(\lm)\nn
&\overset{(a)}{\le}\sum_{y\in\Y}\Gamma(y)\left(-\frac{\zeta^2\lambda^2(x,y)}{2}+\frac{\zeta^2\lambda^{* 2}_\zeta(y)}{2} +\zeta^2\lambda^2(x,y)-\zeta^2\lambda(x,y)\lambda^*_\zeta(y)\right)+\zeta^3K(\lm)\nn
&=\frac{\zeta^2}{2}\sum_{y\in\Y}\Gamma(y)\left(\lambda(x,y)-\lambda^*_\zeta(y)\right)^2+\zeta^3K(\lm)\nn
&=\zeta^2\Psi_{\zeta,x}+\zeta^3K(\lm).
\end{align*}
Here, (a) follows from (\ref{EQ:Lambda_Sum2}), (\ref{EQ:lamba_prime_sum}), and combining all terms involving $\zeta^3$  with $\zeta^3K(\lm)$.

Similarly, one can show that 
$$\E[i^*(X,Y)|X=x]\ge\zeta^2\Psi_{\zeta,x}-\zeta^3K(\lm).$$

Using Taylor's theorem  one can show for all sufficiently small $\zeta$,
\begin{align*}
\left|\left(\log\frac{1+\zeta\lambda(x,y)}{1+\zeta\lambda^*_\zeta(y)}\right)^2-\zeta^2(\lambda(x,y)-\lambda^*_\zeta(y))^2\right|\le \zeta^3 K(\lm).
\end{align*}
Thus,
\begin{align*}
\E\left[(i^*(X,Y))^2|X=x\right]
&=\sum_{y\in\Y}\Gamma(y)\left(1+\zeta\lambda(x,y)\right)\left(\log\frac{1+\zeta\lambda(x,y)}{1+\zeta\lambda^*_\zeta(y)}\right)^2\nn
&\le\sum_{y\in\Y}\Gamma(y)\left(1+\zeta\lambda(x,y)\right)\left(\zeta^2(\lambda(x,y)-\lambda^*_\zeta(y))^2\right)+\zeta^3K(\lm)\nn
&\le\zeta^2\sum_{y\in\Y}\Gamma(y)(\lambda(x,y)-\lambda^*_\zeta(y))^2+\zeta^3K(\lm)\nn
&=2\zeta^2\Psi_{\zeta,x}+\zeta^3K(\lm).
\end{align*}
Hence,
\begin{align*}
\nu_{x,\zeta}&=\text{Var}\left[i^*(X,Y)|X=x \right]\nn
&=\E\left[(i^*(X,Y))^2|X=x\right]-(\E[i^*(X,Y)|X=x])^2\nn
&\le2\zeta^2\Psi_{\zeta,x}+\zeta^3K(\lm).
\end{align*}
 Note that $\E[i^*(X,Y)|X=x]^2\le \zeta^4K(\lm)$.
This gives,
\begin{align*}
\nu_{x,\zeta} &\ge2\zeta^2\Psi_{\zeta,x}-\zeta^3K(\lm).
\end{align*} 

Since  for $x\in\X_\zeta^*$, $\E\left[i^*(X,Y)|X=x\right]=C_\zeta$, for $x\in\X_\zeta^*$ we get
\begin{align*}
\left|\nu_{x,\zeta}-2C_\zeta\right|\le\zeta^3 K(\lm),
\end{align*}
and for $x\notin\X_\zeta^*$,
\begin{align*}
\left|\nu_{x,\zeta}-2(C_\zeta-\rho_{\zeta,x})\right|\le\zeta^3 K(\lm).
\end{align*}
\end{IEEEproof}

Recall that $V_{\text{min},\zeta}$ and $V_{\text{max},\zeta}$  are defined as
\begin{align*}
V_{\text{min},\zeta}\de\min_{P\in\Pi_{W_\zeta}^*}\sum_{x\in\X}P(x)\nu_{x,\zeta}, \\
V_{\text{max},\zeta}\de \max_{P\in\Pi_{W_\zeta}^*}\sum_{x\in\X}P(x)\nu_{x,\zeta},
\end{align*}
where $\Pi_{W_\zeta}^*$ is the set of capacity-achieving probability distributions.

\begin{Lemma}
\label{Le:V_Scaling}
For all sufficiently small $\zeta$, $V_{\text{min},\zeta}$ and $V_{\text{max},\zeta}$  satisfy 
\begin{align*}
\left|V_{\text{min},\zeta}-2C_\zeta\right|\le\zeta^3K(\lm),\nn
\left|V_{\text{min},\zeta}-2\zeta^2\C\right|\le\zeta^3K(\lm)\nn
\left|V_{\text{max},\zeta}-2C_\zeta\right|\le\zeta^3K(\lm),\nn
\left|V_{\text{max},\zeta}-2\zeta^2\C\right|\le\zeta^3K(\lm).
\end{align*}
\end{Lemma}
\begin{IEEEproof}
Note that if $P\in\Pi_{W_\zeta}^*$, then the support of $P$ is contained in $\X_\zeta^*$. Thus from Lemma \ref{Le:IR_Scaling} we get $\left|V_{\text{min},\zeta}-2C_\zeta\right|\le\zeta^3K(\lm)$. 
Moreover since from Lemma \ref{Le:C_Scaling}, $|C_\zeta-\zeta^2\C|\le\zeta^3K(\lm)$, the  inequality $\left|V_{\text{min},\zeta}-2\zeta^2\C\right|\le\zeta^3K(\lm)$ follows. The second set of inequalities for $V_{\text{max},\zeta}$ can be deduced similarly.
\end{IEEEproof}

From  Lemma~\ref{Le:V_Scaling}, we can conclude that $V_{\text{max},\zeta}\approx V_{\text{min},\zeta}$.
Thus taking a hint from Theorem~\ref{Thm:con_vmin_vmax}, we  expect that feedback will not improve the performance of VNCs with respect to the second order coding rate. However, since we have not shown that $V_{\text{max},\zeta}= V_{\text{min},\zeta}$, Theorem~\ref{Thm:con_vmin_vmax} cannot be directly applied here. 
Since $\nu_{x,\zeta}$ is not constant over $x$, even asymptotically,
Theorem~\ref{thm:conv_gen} cannot be applied either.
Thus we prove the converse with a different strategy. 

\label{sec:NA}
Since $\nu_{x,\zeta} \approx 2C_\zeta$ for $x\in\X_\zeta^*$, and for $x\notin\X_\zeta^*$, we have that $\nu_{x,\zeta} \lesssim 2C_\zeta$, to obtain the converse we will add non-negative random variables whenever the input $ X_k\notin\X_\zeta^*$ to ``equalize'' the conditional variance. The following lemma shows the existence of such random variables with desirable properties so that we can apply martingale convergence results. This will yield a proper upper on bound on the maximum possible message set size for sufficiently small $\zeta$.
\begin{Lemma}
\label{Le:zeta}
We can extend   the given probability space to define a sequence of non-negative random variables $\{\xi_{k}\}_{k=1}^n$, such that with  $Z_k=i^*(X_k,Y_k)+\xi_{k}-C_\zeta$,  $\F_k=\sigma(Z_1,\dots,Z_k)$, and for all sufficiently small $\zeta$,
\begin{gather*}
|Z_k| \le 3 \,\quad\mbox{a.s.}, \nn
\E[Z_k|\F_{k-1}]=0\quad\mbox{a.s.}, \nn
V_{\text{min},\zeta}-\zeta^3K(\lm)\le\E[Z^2_k]\le V_{\text{min},\zeta}+\zeta^3K(\lm),\nn
V_{\text{max},\zeta}-\zeta^3K(\lm)\le\E[Z^2_k]\le V_{\text{max},\zeta}+\zeta^3K(\lm),\nn
\left\| \frac{\sum_{k=1}^n\E[Z^2_k|\F_{k-1}]}{\sum_{k=1}^n\E[Z^2_k]}-1\right\|_\infty^{1/2}\le \sqrt{\zeta}K(\lm).
\end{gather*}
\end{Lemma}
\begin{IEEEproof}
For each $x\notin\X_\zeta^*$, define $\{\xi_{x,k}\}_{k=1}^n$ to be a sequence of i.i.d. random variables, independent of all other random variables such that 
$$P(\xi_{x,k}=\rho_{\zeta,x}+2)=1-P(\xi_{x,k}=0)=\frac{\rho_{\zeta,x}}{\rho_{\zeta,x}+2}.$$ 
The variance of the above random variable is
\begin{align*}
\text{Var}[\xi_{x,k}]
&=\E[(\xi_{x,k})^2]-\left(\E[\xi_{x,k}]\right)^2\nn
&=\frac{\rho_{\zeta,x}}{(\rho_{\zeta,x}+2)}(\rho_{\zeta,x}+2)^2-\left(\frac{\rho_{\zeta,x}}{(\rho_{\zeta,x}+2)}(\rho_{\zeta,x}+2)\right)^2\nn
&=2\rho_{\zeta,x}.
\end{align*}
Let 
$$\xi_{k}=\sum_{x\notin\X_\zeta^*}\xi_{x,k}\mathbf{1}\{X_k=x\}.$$
Then,
\begin{align*}
|Z_k|&\le |i^*(X_k,Y_k)|+\xi_{k}+ C_\zeta\nn
 &\le |i^*(X_k,Y_k)|+ \max_{x\notin\X_\zeta^*}\rho_{\zeta,x}+2+C_\zeta\nn
 &\le 3 \quad \mbox{a.s.},
\end{align*} 
for all sufficiently small $\zeta$.
Let $\G_k=\sigma(M,Y_1,\xi_1,\dots,Y_k,\xi_k)$. We note that $X_{k}$ is $\G_{k-1}$ measurable (since the  message $M$ and  past outputs $(Y_1,\dots, Y_{k-1})$ determine the input $X_k$) and $\F_k\subseteq\G_k$.
Thus,
\begin{align*}
\E[i^*(X_k,Y_k)|\G_{k-1}]&=\E[i^*(X_k,Y_k)|X_{k}]\nn
&=C_\zeta-\rho_{\zeta,X_k}\mathbf{1}\{X_k\notin\X_\zeta^*\}.
\end{align*}

Then,
\begin{align*}
\E[Z_k|\G_{k-1}]
&=C_\zeta-\rho_{\zeta,X_k}\mathbf{1}\{X_k\notin\X_\zeta^*\}+\rho_{\zeta,X_k}\mathbf{1}\{X_k\notin\X_\zeta^*\}-C_\zeta\\
&=0.
\end{align*}
Taking the conditional expectation with respect to $\F_{k-1}$, and since  $\F_{k-1}\subseteq\G_{k-1}$,
\begin{align*}
\E[Z_k|\F_{k-1}]=0.
\end{align*}
Also,
\begin{align}
\E[Z^2_k|\G_{k-1}]
=&\V[Z_k|\G_{k-1}]\nn
\overset{}{=}&\V[i^*(X_k,Y_k)+\xi_{k}|\G_{k-1}]\nn
\overset{(a)}{=}&\V[i^*(X_k,Y_k)|\G_{k-1}]+\V[\xi_{k}|\G_{k-1}]\nn
\overset{(b)}{\le}& \,2C_\zeta-2\rho_{\zeta,X_k}\mathbf{1}\{X_k\notin\X_\zeta^*\}+\zeta^3K(\lm)+2\rho_{\zeta,X_k}\mathbf{1}\{X_k\notin\X_\zeta^*\}\nn
=&\,2C_\zeta+\zeta^3K(\lm).
\label{EQ:ZCU}
\end{align}
Here (a) follows since given $X_k$, $i^*(X_k,Y_k)$ and $ \xi_{k}$ are conditionally independent, and\\* 
(b) follows from Lemma \ref{Le:IR_Scaling} and noting that $\text{Var}[\xi_{x,k}]=2\rho_{\zeta,x}$.

Similarly
\begin{align}
\E[Z^2_k|\G_{k-1}]\ge2C_\zeta-\zeta^3K(\lm).
\label{EQ:ZCL}
\end{align}
Thus from Lemma \ref{Le:V_Scaling}, (\ref{EQ:ZCU}) and (\ref{EQ:ZCL}),
\begin{align*}
V_{\text{min},\zeta}-\zeta^3K(\lm)\le\E[Z^2_k|\G_{k-1}]\le V_{\text{min},\zeta}+\zeta^3K(\lm),\\
V_{\text{max},\zeta}-\zeta^3K(\lm)\le\E[Z^2_k|\G_{k-1}]\le V_{\text{max},\zeta}+\zeta^3K(\lm).
\end{align*}
Once again taking the conditional expectation with respect to $\F_{k-1}$,
\begin{align}
V_{\text{min},\zeta}-\zeta^3K(\lm)\le\E[Z^2_k|\F_{k-1}]\le V_{\text{min},\zeta}+\zeta^3K(\lm),\label{EQ:Var_Fk}\\
V_{\text{max},\zeta}-\zeta^3K(\lm)\le\E[Z^2_k|\F_{k-1}]\le V_{\text{max},\zeta}+\zeta^3K(\lm).\nonumber
\end{align}

Now consider
\begin{align*}
\frac{\sum_{k=1}^n\E[Z^2_k|\F_{k-1}]}{\sum_{k=1}^n\E[Z^2_k]}-1 &\le \frac{V_{\text{min},\zeta}+\zeta^3K(\lm)}{V_{\text{min},\zeta}-\zeta^3K(\lm)}-1 \\
&=\frac{2\zeta^3K(\lm)}{V_{\text{min},\zeta}-\zeta^3K(\lm)}.
\end{align*}
Here, we note that for the last equality to hold, the constants $K(\lm)$ appearing in the left and right terms of (\ref{EQ:Var_Fk}) should be equal. If they are not, we simply replace each by  the maximum of the two constants.
Similarly,
\begin{align*}
\frac{\sum_{k=1}^n\E[Z^2_k|\F_{k-1}]}{\sum_{k=1}^n\E[Z^2_k]}-1 \ge -\frac{2\zeta^3K(\lm)}{V_{\text{min},\zeta}+\zeta^3K(\lm)}.
\end{align*}
Thus,
\begin{align*}
\left\| \frac{\sum_{k=1}^n\E[Z^2_k|\F_{k-1}]}{\sum_{k=1}^n\E[Z^2_k]}-1\right\|_\infty^{1/2} &\le \left(\frac{2\zeta^3K(\lm)}{V_{\text{min},\zeta}-\zeta^3K(\lm)}\right)^{1/2}\nn
&\le \sqrt{\zeta}K(\lm),
\end{align*}
where the last inequality is due to Lemma \ref{Le:V_Scaling}.
\end{IEEEproof}

Now we give the proof of Theorem~\ref{Thm:conv_vnc}.
Define
\begin{align}
\kappa_{\zeta,n}&\de K(\lm)\left(\frac{\log(n)}{\zeta^3\sqrt{n}}+\sqrt{\zeta}\right)\\
r_{\zeta,n} &\de 
\begin{cases}
\sqrt{\left(V_{\text{min},\zeta}-\zeta^3 K(\lm)\right)}\Phi^{-1}(\varepsilon+\kappa_{\zeta,n}), &0<\varepsilon\le\frac{1}{2}-\kappa_{\zeta,n}\\
\sqrt{\left(V_{\text{max},\zeta}+\zeta^3 K(\lm)\right)}\Phi^{-1}(\varepsilon+\kappa_{\zeta,n}), & \frac{1}{2}-\kappa_{\zeta,n} < \varepsilon < 1.
\end{cases}
\label{EQ:r_n_vnc}\\
\rho_{\zeta,n} &\de\exp(nC_\zeta + \sqrt{n}r_n),
\end{align}
Now defining $\{\xi_k\}_{k=1}^n$ as a sequence of random variables as in Lemma \ref{Le:zeta}, consider the following chain of inequalities: 
\begin{align}
P\left(\sum_{k=1}^n i^*(X_k,Y_k)\ge \log \rho_{\zeta,n} \right)&\overset{(a)}{\le} P\left(\sum_{k=1}^n i^*(X_k,Y_k)+\xi_k \ge \log \rho_{\zeta,n} \right)\nn
   &\overset{(b)}{=}P\left(\sum_{k=1}^n Z_k \ge  \sqrt{n}r_{\zeta,n} \right)\nn
&\overset{(c)}{\le}P\left(\frac{1}{\sqrt{\sum_{k=1}^n \E[Z^2_k]}}\sum_{k=1}^n Z_k \ge \Phi^{-1}(\varepsilon+2\kappa_{\zeta,n}) \right)\nn
&\overset{(d)}{\le} 1-\varepsilon-2\kappa_{\zeta,n}+\chi \cdot \left(\frac{n\log(n)}{\left(\sum_{k=1}^n \E[Z^2_k]\right)^{3/2}}+\left\| \frac{\sum_{k=1}^n\E[Z^2_k|\F_{k-1}]}{\sum_{k=1}^n\E[Z^2_k]}-1\right\|_\infty^{1/2}\right)\nn
&\overset{(e)}{\le} 1- \varepsilon-2\kappa_{\zeta,n}+\left(K(\lm)\frac{\log(n)}{\zeta^3\sqrt{n}}+K(\lm)\sqrt{\zeta}\right) \nn
&=1- \varepsilon-\kappa_{\zeta,n}.
\end{align}
Here,
(a)  follows since $\xi_k$ is a non-negative random variable, \\
(b) follows from setting $Z_k$ as in Lemma \ref{Le:zeta},\\
(c) follows since $n(V_{\text{min},\zeta}-\zeta^3K(\lm))\le\sum_{k=1}^n \E[Z^2_k]\le n(V_{\text{max},\zeta}+\zeta^3K(\lm))$ due to Lemma \ref{Le:zeta},\\
(d) follows from the martingale central limit theorem \cite[Corollary to Theorem 2]{Bolthausen}, and taking the constant as $\chi$ (which does not depend upon the channel or $n$), and \\
(e) follows from noting that $\left(\sum_{k=1}^n \E[Z^2_k]\right)^{3/2} \ge n\sqrt{n}(2\zeta^2\C-\zeta^3K(\lm))^{3/2}$, and then absorbing $\chi$ into  $K(\lm)$.

Invoking Lemma~\ref{thm:gen_conv_code} from the Appendix with $q_\zeta(\by^n)=\prod_{i=1}^n q_\zeta^\ast(y_i)$, we get
\begin{align*}
    \log M^\ast_{\textnormal{fb},\zeta}(n, \varepsilon) \le \log \rho_{\zeta,n}- \log \kappa_{\zeta,n} \le nC_\zeta + \sqrt{n}r_{\zeta,n} -\log \kappa_{\zeta,n}.
\end{align*}

If $0<\varepsilon < \frac{1}{2}$,
\begin{align*}
    \limsup_{n\to\infty}\frac{\log M^\ast_{\textnormal{fb},\zeta}(n, \varepsilon) -nC_\zeta}{\sqrt{nV_{\text{min},\zeta}}} \le \sqrt{1-\frac{\zeta^3 K(\lm)}{V_{\text{min},\zeta}}}\Phi^{-1}(\varepsilon+K(\lm)\sqrt{\zeta}),
\end{align*}
and hence,
\begin{align*}
    \limsup_{\zeta\to 0}\limsup_{n\to\infty}\frac{\log M^\ast_{\textnormal{fb},\zeta}(n, \varepsilon)-nC_\zeta}{\sqrt{nV_{\text{min},\zeta}}}\le \Phi^{-1}(\varepsilon).
\end{align*}
Similarly, when $\frac{1}{2} \le \varepsilon<1$,
\begin{align*}
    \limsup_{\zeta\to 0}\limsup_{n\to\infty}\frac{\log M^\ast_{\textnormal{fb},\zeta}(n, \varepsilon) -nC_\zeta}{\sqrt{nV_{\text{max},\zeta}}}\le \Phi^{-1}(\varepsilon).
\end{align*}
Since $V_{\text{min},\zeta}/V_{\text{max},\zeta} \rightarrow 1$ as
$\zeta \rightarrow 0$ by Lemma~\ref{Le:V_Scaling}, the conclusion follows.
\hfill\IEEEQED 

\section*{Acknowledgment}
This research was supported by
the National Science Foundation under grant CCF-1513858.

\bibliographystyle{IEEEtran}
\bibliography{IEEEabrv,DMC_FB}

\appendix

As noted in the introduction, the problem of maximizing the
second-order coding rate with feedback is related to
the design of controlled random walks.
\begin{definition}
A \emph{controller} is a function $F:(\X \times \Y)^\ast \rightarrow \PP(\X)$.
\end{definition}
We shall sometimes write $F(\cdot|\bx^k,\by^k)$ for $F(\bx^k,\by^k)(\cdot)$.
Given a controller $F$,
let $F \circ W$ denote the distribution
\begin{equation}
(F \circ W)(\bx^n,\by^n) = \prod_{k = 1}^n F(x_k|\bx^{k-1},\by^{k-1}) W(y_k|x_k)
\end{equation}
and let $FW(\by^n)$ denote the marginal over $\bY^n$ induced by
$F \circ W$.

The following lemma shows that any controller gives rise to an
achievable second-order coding rate. The idea is to use the controller
to generate a random ensemble of feedback codes and then use
a technique that dates back to Shannon~\cite{shannon57} to
bound the error probability of this ensemble.

\begin{Lemma}[Achievability]
\label{lemma:shannon:achievability:FB}
For any controller $F$ and any $n$, $\theta$, and rate $R$,
\begin{equation}
\label{eq:Shannon:FB:achievability}
\bar{\mP}_{\me, \textnormal{fb}}(n,R)
 \le (F \circ W)\left(\frac{1}{n} \log
                     \frac{W(\bY^n|\bX^n)}{FW(\bY^n)} \le R + 
                      \theta\right) + e^{-n \theta}.
\end{equation}
Thus, if for some $\alpha$ and $\varepsilon$,
\begin{align}
\label{eq:achievability:control:hypothesis}
\limsup_{n\to\infty}\inf_F(F\circ W)\left(\sum_{k=1}^n \log \frac{W(Y_k|X_k)}{(F W)(Y_k|\bY^{k-1})}\le nC + \alpha\sqrt{n} \right) < \varepsilon,
\end{align}
then
\begin{align}
\label{eq:achievability:control}
\liminf_{n\to\infty}\frac{\log M^\ast_{\textnormal{fb}}(n, \varepsilon)-n C}{\sqrt{n }} \geq \alpha.
\end{align}
\end{Lemma}
\begin{IEEEproof}
We begin by showing~(\ref{eq:Shannon:FB:achievability}).
Consider a random code in which, for each message, the channel
input at time $k$ when the past inputs are $\bx^{k-1}$ and the
past outputs are $\by^{k-1}$ is chosen according to
$F(\cdot|\bx^{k-1},\by^{k-1})$. That is, $f(m,\by^{k-1})$ is chosen
randomly according to
\begin{equation}
F(\cdot|(f(m,\emptyset),f(m,y_1),\ldots,f(m,\by^{k-2})), \by^{k-1}).
\end{equation}
Given $\by^n$, the decoder selects the message with the lowest 
index that achieves the minimum over $m$ of
\begin{equation}
\prod_{k = 1}^n W(y_k|f(m,\by^{k-1})).
\end{equation}

By the union bound and other standard steps, the ensemble average error
probability of this code is upper bounded by
\begin{align}
&  \sum_{\bx^n,\by^n} (F \circ W) (\bx^n,\by^n) \b1{\left\{ \frac{1}{n} \log
     \frac{W(\by^n|\bx^n)}{FW(\by^n)} \le R + \theta \right\} } \\
   & \phantom{\sum}  + e^{nR} \sum_{\bx^n,\by^n} (F \circ W) (\bx^n,\by^n) 
      \sum_{\tilde{\bx}^n: W(\by^n|\tilde{\bx}^n) \ge W(\by^n|\bx^n)}
         \prod_{k = 1}^n F(\tilde{x}_k|\tilde{\bx}^{k-1},\by^{k-1})
          \b1\left\{\frac{1}{n} \log 
                \frac{W(\by^n|\bx^n)}{FW(\by^n)} > R + \theta\right\} \\
   & \le (F \circ W)\left(\frac{1}{n} \log \frac{W(\bY^n|\bX^n)}{FW(\bY^n)}
                   \le R + \theta\right) \\
   &  \phantom{\sum}   + e^{nR} \sum_{\by^n} FW(\by^n)
          \sum_{\tilde{\bx}^n}
         \prod_{k = 1}^n F(\tilde{x}_k|\tilde{\bx}^{k-1},\by^{k-1})
          \b1\left\{\frac{1}{n} \log 
          \frac{W(\by^n|\tilde{\bx}^n)}{FW(\by^n)} > R + \theta\right\} \\
    & \le (F \circ W)\left(\frac{1}{n} \log \frac{W(\bY^n|\bX^n)}{F
               W(\bY^n)} \le R + \theta\right) \\
    & \phantom{\sum}   + e^{nR} e^{-n(R + \theta)} \sum_{\tilde{\bx}^n} 
               \sum_{\by^n} 
         \prod_{k = 1}^n F(\tilde{x}_k|\tilde{\bx}^{k-1},\by^{k-1})
                   W(y_k|\tilde{x}_k),
\end{align}
which implies~(\ref{eq:Shannon:FB:achievability}). 
Now
suppose~(\ref{eq:achievability:control:hypothesis}) holds and in 
(\ref{eq:Shannon:FB:achievability}), select $R = C + \alpha^\prime/\sqrt{n}$
and $\theta = n^{- \beta}$ for some $1/2 < \beta < 1$ and
$\alpha^\prime < \alpha$. Then we have
\begin{equation}
\limsup_{n \rightarrow \infty} \bar{\mP}_{\me, \textnormal{fb}}\left(n,
C + \frac{\alpha^\prime}{\sqrt{n}}\right) \le \limsup_{n \rightarrow \infty}
   \inf_F (F \circ W) \left(\frac{1}{n} \log
     \frac{W(\bY^n|\bX^n)}{FW(\bY^n)} \le C + \frac{\alpha^\prime}{\sqrt{n}}
                  + \frac{1}{n^\beta} \right).
\end{equation}
Thus if (\ref{eq:achievability:control:hypothesis}) holds we have
\begin{equation}
\limsup_{n \rightarrow \infty} \bar{\mP}_{\me, \textnormal{fb}}\left(n,
C + \frac{\alpha^\prime}{\sqrt{n}}\right) < \varepsilon,
\end{equation}
since $\alpha^\prime < \alpha$.
This implies that eventually, 
\begin{equation}
\log M^\ast_{\textnormal{fb}}(n, \varepsilon) \ge n C + \alpha^\prime \sqrt{n}.
\end{equation}
Since this holds for any $\alpha^\prime < \alpha$, 
(\ref{eq:achievability:control}) follows.
\end{IEEEproof}

The next result is used repeatedly in the paper as a starting point in proving converses. A similar inequality to (\ref{eq:metaconverse}) can be found in~\cite[(42)]{Fong-Tan}. Observe that 
(\ref{eq:controlled:converse:hyp}) and
(\ref{eq:controlled:converse:conc}), which 
are a consequence of (\ref{eq:metaconverse}), are nearly a converse 
of (\ref{eq:achievability:control:hypothesis}) and 
(\ref{eq:achievability:control}) above.

\begin{Lemma}[Converse]
For any $\rho>0$ and $\varepsilon > 0$
\label{thm:gen_conv_code}
\begin{align}
\label{eq:metaconverse}
\log M_{\textnormal{fb}}^\ast(n, \varepsilon) \le \sup_{F} \inf_{q\in\PP(\Y^n)}\left(\log \rho-\log \left[\left(1-\varepsilon - (F\circ W)\left(\sum_{k=1}^n\log \frac{ W(Y_k|X_k)}{q(Y_k|\bY^{k-1})} > \log \rho \right)\right)^+\right]\right).
\end{align}
In particular, if for some $\alpha$ and $\varepsilon$,
\begin{align}
\label{eq:controlled:converse:hyp}
\liminf_{n\to\infty}\inf_F\sup_{q\in\PP(\Y^n)}(F\circ W)\left(\sum_{k=1}^n\log \frac{W(Y_k|X_k)}{q(Y_k|\bY^{k-1})}\le nC + \alpha\sqrt{n} \right)>\varepsilon,
\end{align}
then 
\begin{align}
\label{eq:controlled:converse:conc}
\limsup_{n\to\infty}\frac{\log M^\ast_{\textnormal{fb}}(n, \varepsilon)-n C}{\sqrt{n }} \le \alpha.
\end{align}
\end{Lemma}
\begin{IEEEproof}
Consider an $(n,R)$ feedback code $(f,g)$ with average error probability
at most $\varepsilon$. We will denote this code by $\cd$
and its average error probability by $\varepsilon_\cd$.
Define
\begin{equation*}
M_{\textnormal{fb},\cd}^\ast(n) \de \lceil \exp\left({nR}\right) \rceil.  
\end{equation*}
Then
\begin{align*}
M_{\textnormal{fb}}^\ast(n,\varepsilon)= \sup_{\cd:\varepsilon_\cd \le \varepsilon} M_{\textnormal{fb},\cd}^\ast(n).
\end{align*}

The code $\cd$ induces a controller $F$ via
\begin{align*}
F(x_k|\bx^{k-1},\by^{k-1})\de \frac{1}{M_{\textnormal{fb},\cd}^\ast(n)}\sum_{m=1}^{M_{\textnormal{fb},\cd}^\ast(n)}\1\{f(m,\by^{k-1})=x_k\},
\end{align*}
which, in fact, does not depend on $\bx^{k-1}$.
Now consider the problem of hypothesis testing where a random variable $U$ taking values in $\cU$ can have probability measure $P$ or $Q$. Upon observing $U$, the goal is to declare either $U\sim P$ (hypothesis $H_1$) or $U\sim Q$ (hypothesis $H_2$). Let $\beta_\alpha(P,Q)$ denote the minimum attainable  error probability under $Q$ when the  error probability under $P$ does not exceed $1-\alpha$. Then the Neyman-Pearson lemma~\cite[Proposition II.D.1, p. 33]{Poor94} guarantees that there exists a (possibly randomized) test   
 $T:\cU\to\{0,1\}$ (where $0$ corresponds to the test selecting $Q$) such that
 \[
 \sum_{u\in\cU}P(u)T(1|u)\ge\alpha,\nn
 \sum_{u\in\cU}Q(u)T(1|u)=\beta_\alpha(P,Q).
 \]
Then for any $\rho>0$
\begin{align}
\alpha-\rho\beta_\alpha(P,Q)&\le \sum_{u\in\cU}T(1|u)(P(u)-\rho Q(u))\nn
&\le \sum_{u\in\cU}T(1|u)(P(u)-\rho Q(u)) \1\{P(u)>\rho Q(u)\}\nn
&=P\left(\frac{P(u)}{Q(u)}>\rho, T=1 \right)-\rho Q\left(\frac{P(u)}{Q(u)}>\rho, T=1 \right)\nn
&\le P\left(\frac{P(u)}{Q(u)}>\rho\right).
\label{eq:HT_conv}
\end{align}

Fix a $q\in\PP(\Y^n)$. Applying~\cite[Theorem 26]{Polyanskiy} (with $Q_{Y|X}=q$, $\varepsilon^\prime=1-1/M_{\textnormal{fb},\cd}^\ast(n)$;
the assertion there is without feedback but one can verify that it
applies to the feedback case as well), we get
\begin{align*}
\beta_{1-{\varepsilon_\cd}}\left(F\circ W,F\circ q\right)\le\frac{1}{M_{\textnormal{fb},\cd}^\ast(n)}.
\end{align*}
Moreover, from~(\ref{eq:HT_conv})
\begin{align*}
\alpha\le (F\circ W)\left(\frac{d(F\circ W)}{d(F\circ q)} > \rho \right)+\rho \beta_{\alpha}\left(F\circ W,F\circ q\right),
\end{align*}
i.e.,
\begin{align*}
\beta_{1-{\varepsilon_\cd}}\left(F\circ W,F\circ q\right)\ge \frac{1}{\rho}\left(1-{\varepsilon_\cd}- (F\circ W)\left(\frac{d(F\circ W)}{d(F\circ q)} > \rho \right)\right)^+.
\end{align*}
Thus
\begin{align*}
\log M_{\textnormal{fb},\cd}^\ast(n) \le \log \rho - \log \left[\left(1-{\varepsilon_\cd}- (F\circ W)\left(\frac{d(F\circ W)}{d(F\circ q)} > \rho \right)\right)^+\right].
\end{align*}
Using the fact that $\varepsilon_\cd \le \varepsilon$ and that $q$ was
arbitrary, we obtain
\begin{align*}
\log M_{\textnormal{fb},\cd}^\ast(n) \le \inf_{q\in\PP(\Y^n)} \log \rho - \log \left[\left(1-{\varepsilon}- (F\circ W)\left(\frac{d(F\circ W)}{d(F\circ q)} > \rho \right)\right)^+\right].
\end{align*}
Taking the supremum over all controllers $F$ and noting that
\begin{equation*}
 \frac{d(F\circ W)}{d(F\circ q)}=\prod_{k=1}^n \frac{W(y_k|x_k)}{q(y_k|\by^{k-1})},
\end{equation*} we get
\begin{align*}
\log M_{\textnormal{fb}}^\ast(n) \le \sup_{F} \inf_{q\in\PP(\Y^n)} \left(\log \rho-\log \left[\left(1-\varepsilon - (F\circ W)\left(\sum_{k=1}^n\log \frac{ W(Y_k|X_k)}{q(Y_k|\bY^{k-1})} > \log \rho \right)\right)^+\right]\right).
\end{align*}
This establishes (\ref{eq:metaconverse}). (\ref{eq:controlled:converse:conc}) 
follows directly from (\ref{eq:metaconverse}) and 
(\ref{eq:controlled:converse:hyp}).
\end{IEEEproof}
\end{document}